\documentstyle[11pt,aaspp4]{article}

\slugcomment{To appear in January 1999 A.J.}

\lefthead{GIZIS \& REID}
\righthead{M SUBDWARFS}

\begin{document}

\title{M SUBDWARFS:  THE POPULATION II LUMINOSITY FUNCTION}

\author{\sc John E. Gizis\altaffilmark{1} and I. Neill Reid\altaffilmark{2}}
\affil{Palomar Observatory, 105-24, California Institute of Technology,
Pasadena, California 91125\\ e-mail: gizis@stratford.phast.umass.edu, inr@astro.caltech.edu}

\altaffiltext{1}{Current Address: Department of Physics and
Astronomy, LGRT 532A, University of Massachusetts, Amherst MA 01003-4525}
 
\altaffiltext{2}{Visiting Research Associate, Carnegie Institute of Washington}

\begin{abstract}
We present results of a study of very low mass halo stars.  
Using a sample of proper motion stars identified from plate material
taken as part of the first and second Palomar Sky Surveys, we measure 
the space density, metallicity distribution, and kinematics of the 
Population II M subdwarfs. Our overall luminosity function is
in good agreement with previous analyses of the space density of
nearby very-low-mass halo subdwarfs, and confirms the discrepancy
between local analyses and the space densities inferred from 
deep HST starcounts.
We show for the first time that both the metallicity
distribution and kinematics of late-type halo subdwarfs are consistent with
those of their higher mass metal-poor counterparts. Dividing our sample
by abundance, we find no evidence that the mass function of field halo
stars is dependent upon metallicity.  We provide data for three
nearby subdwarfs that may merit additional observations.
\end{abstract}


\section{Introduction \label{halolf-intro}}

Long-lived low-mass stars are a local relic of the earliest
stages of the Galaxy's formation.  Dubbed the Population II in
contrast to the disk Population I, these stars make up the Galactic 
halo (sometimes called spheroid).  Although comprising only a very small
faction of the stars in the Galaxy, Population II stars provide
one of the most important records of Galactic history.  

The Population II luminosity function is of particular interest
since its derivation is a requirement in most studies of the stellar 
mass function. Determination of the stellar 
mass function as function of Galactic epoch and metallicity provides 
significant constraints upon the theory of star formation.  
Evolution of the mass function with time
or metallicity might be expected since conditions in
star-forming clouds were likely much different in the protoGalaxy
(c.f. \cite{z95}).  Moreover, since halo brown dwarfs have faded far below
the threshold for even near-infrared detection, the form of the 
subdwarf mass function near the hydrogen-burning limit represents
the most effective technique (in the absence of gravitational
lensing detections) for estimating the Population II dark matter contribution.

Despite the intuitive simplicity of the process of counting
stars to determine the luminosity function, there have 
substantial disagreements between different studies in
both the normalization and slope of the luminosity function
(see the review of \cite{mould96}).  Two recent surveys have
measured the space density of very-low-mass Population II stars, 
the M subdwarfs.
Counts of faint, distant M subdwarfs using HST
(\cite{gfb98}) predict $\sim 2.5$ times fewer stars at
$M_V=11$ than are found locally by Dahn {et al.} (1995).
Dahn {et al.} (1995) base their study upon a sample of
the local high proper motion stars in the LHS Catalog (\cite{lhs})
for which they have measured accurate trigonometric parallaxes.
 
In this paper, we determine the luminosity function of the field
Population II M subdwarfs using a new proper motion survey based upon 
plates taken as part of the
first and second Palomar Sky Surveys.  In Section~\ref{general:halolf},
we discuss some general problems with the selection of halo stars.
In Section~\ref{data:halolf}, we 
describe the data used in our survey and the methods used to select a
halo sample.  In Section~\ref{lf:halolf}, we use our data to measure the
M subdwarf luminosity function.
In Section~\ref{discussion:halolf}, we compare our luminosity function
to other published luminosity functions and discuss its significance.  
In Section~\ref{conclusion:halolf}, we summarize our results.
In the Appendix, we describe some stars that may be of particular
interest.  

\section{Selection of Halo Stars --- General Principles\label{general:halolf}}

Measurement of the Population II luminosity function is 
considerable more difficult than the measurement of the
Population I luminosity function, for the simple reason that
the Population II represents a tiny fraction of stars locally.
One must either count very faint stars at large distances from the
Galactic plane (\cite{gfb98}), in which case one sacrifices
the ability to obtain spectroscopic data, or else determine some
way to identify the local Population II stars by excluding
the local $\sim 99.8\%$ of stars that belong to the disk.   The
latter course requires an efficient method of eliminating the disk stars
before time-consuming followup observations are obtained. As members of
a high velocity-dispersion, low rotation population, subdwarfs tend to
have high heliocentric velocities, and choosing
an initial sample of proper motion stars increases the halo
contribution from $\sim 0.2\%$ to $\sim 30\%$ (Schmidt 1975).
Such selection criteria introduce some kinematic bias, but lead to a much more 
manageable sample.  Nevertheless, such a sample is still dominated by 
old disk (Population I) and ``thick disk'' (Intermediate Population II, IPII) 
stars\footnote{The term Intermediate Population II, used instead of
``thick disk'' or ``extended disk,'' should not be confused
with the Population II halo.  Considerable evidence exists that
the two populations are kinematically and chemically discrete 
(\cite{majewski}).}.

As recognized by Schmidt (1975) and further discussed
by Bahcall \& Casertano (1986, hereafter BC), even a small proportion of 
contamination by high velocity disk stars can
lead a gross overestimate of the space density of the halo.
In order to ensure a pure halo sample for his study, Schmidt imposed
a restrictive velocity criterion ($v_{tan} > 250$ km/s)
and then applied a correction to account for halo stars
with smaller velocities.   BC used Monte Carlo
simulations to estimate a correction factor of 3.03 for a cutoff
velocity of $220$ km s$^{-1}$ (this correction is dependent on the kinematics
assumed for the underlying population). Most recently, this technique has
been used to derive a halo luminosity function from 
the LHS catalog stars with 
$0.8 \arcsec {\rm yr}^{-1} \le \mu \le 2.50 \arcsec {\rm yr}^{-1}$ and
$11.0 < m_r < 18.1$ (\cite{dlhg94}).

If trigonometric parallaxes are available for the target stars,
then the distances and tangential velocities are estimated easily
and the appropriate kinematic criteria can be applied.  This is
the case for the Schmidt (1975) sample and the Dahn {et al.} (1995)
samples.   Measurement of a useful trigonometric parallax requires
years of observing --- and the  
the target must be relatively close ($d \lesssim 100$ pc).  
Expansion of the sample to 
cover a larger space volume requires a less direct method of distance 
estimation.  Photometric parallaxes provide that method, with the calibrating
color-absolute magnitude relations provided by subdwarfs with high quality 
trigonometric parallaxes.  

\begin{figure}
\plotone{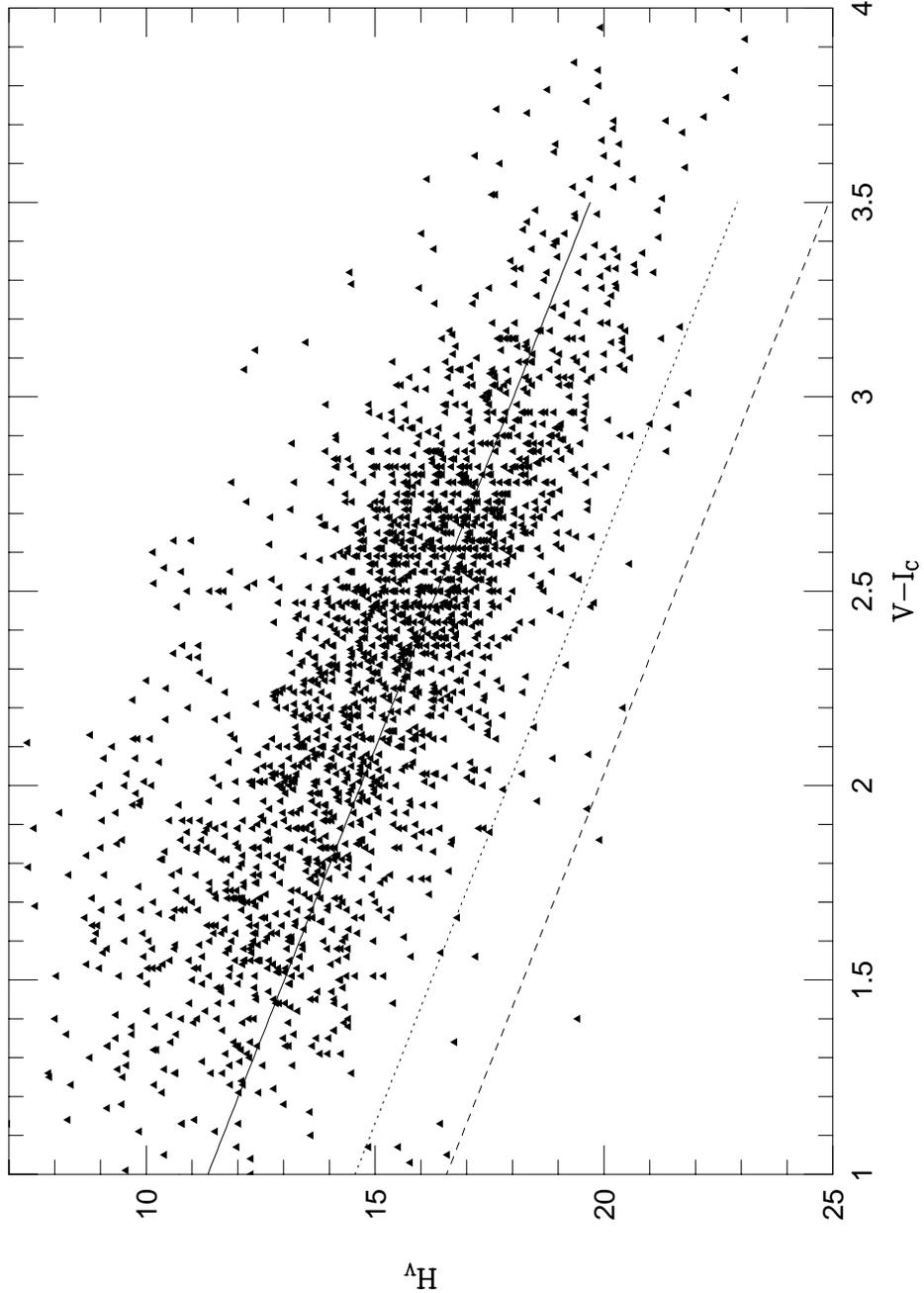}
\caption{A color - reduced proper motion ($V-I,H_V$) 
diagram for a nearly volume limited sample of stars within
25 parsecs.  The 2111 observable known nearby 
M dwarfs (Reid {et al.} 1995; Hawley {et al.} 1996) 
are plotted.  For some stars, the V-I photometry was estimated on the
basis of their spectral type.  The solid line plots a disk population
using the Stobie {et al.} (1989) main sequence and $v_{tan} = 50$ km s$^{-1}$ .
The dotted line plots the same main sequence with $v_{tan} = 220$ 
km s$^{-1}$.  The dashed line plots a main
sequence that is two magnitudes subluminous, roughly corresponding to 
the sdM.  It is clear that volume limited samples are a poor way of
selecting Population II samples. }
\label{figure-rpm-demo}
\end{figure}

The efficiency of the initial selection of halo candidates
can be increased by use of the ``reduced proper motion'', H, first  
used by Luyten (1939).  It is defined as 
\begin{equation}
H =  m + 5\log \mu +5 = M + 5 \log {{v_{tan}}\over{4.74}}
\end{equation}
where m is the apparent magnitude and M is the absolute magnitude.
Note that $H$ is distance independent and determined entirely by
observables.  If we assume a color-magnitude relationship, we
can plot constant tangential velocity contours on the data.  
Figure~\ref{figure-rpm-demo} illustrates the concept, where data are plotted
for 2111 M dwarfs within $\sim$30 parsecs of the Sun(Reid {et al.} 1995;
Hawley {et al.} 1996).
(Note that the increasing proper motion bias for redder stars
in the Gliese \& Jahreiss 1991 nearby star catalog is clearly evident.)  
The lines illustrate 
where a disk population with halo-like velocities will lie, 
and how halo stars will be even more easily distinguished
in the diagram.   

As subluminous stars in the (M$_I$, (R-I)) and (M$_I$, (V-I)) planes, 
halo stars have systematically larger H than disk dwarfs at a given 
color, mimicking the effects of increasing $v_{tan}$. In contrast, the
coolest metal-poor subdwarfs are superluminous in the (M$_V$, (B-V)) plane
(Gizis 1997), and a cutoff in the (H$_V$, (B-V)) diagram tends to
discriminate against lower-abundance stars. We have therefore
defined our sample of candidate halo subdwarfs as those stars with reduced
proper motion, H$_I$, which exceed the value predicted for {\sl disk}
dwarfs with $v_{tan}=220$km s$^{-1}$ and of the appropriate (V-I) or
(R-I) color. Since the halo stars are subluminous, this selection corresponds 
to a cutoff velocity that is a function of metallicity - an effect which
is taken into account in the subsequent analysis.

\section{Selection of Halo Stars --- Data and Method\label{data:halolf}}


We utilized the techniques described above to define our sample.
Proper motion stars were identified using the first and second
epoch 48-inch Schmidt photographic plates, as 
described in (Section~\ref{mumeasure}).
An initial cut of the sample was made
to select large tangential velocity stars (Section~\ref{rpm}).   
We then obtained
spectra of each candidate halo star to estimate its 
metallicity and radial velocity
(Section~\ref{spectroscopy:halolf}).  

\subsection{Proper Motions\label{mumeasure}}

Table~\ref{table-plates} lists the fields analyzed in the current survey. 
In each, the
effective area is $\sim25$ square degrees, with most having plate material
spanning a baseline of at least 40 years. In each case, the first epoch 
plates were
taken with the 48-in. Oschin Telescope as part of
the first Palomar Observatory Sky Survey (hereafter POSSI).  
For the most part, the second epoch plates were taken in 
the 1980s by either the
48-in. Oschin Telescope (in the course of the 
Palomar Observatory Second Sky Survey, POSSII) or by the
U.K. Schmidt Telescope (UKST).  These second epoch plates
were used by Tinney, Reid, \& Mould (1993, hereafter TRM)
in their determination of the disk luminosity function. 
All of the POSSI plates were scanned by the APM facility 
at the Institute of Astronomy, Cambridge, while the
POSSII/UKST plates were scanned by COSMOS at the Royal Observatory, Edinburgh.
One additional field, centered near the North Galactic Pole, was included.
As described by Reid (1990), the second epoch plates for this field were 
taken at Palomar in 1976, and all plate material was scanned by COSMOS. Each of
these scans yields positions (both pixel x,y positions on the plate
and right ascension, declination positions on the sky), magnitudes,
and morphological classification as stars, galaxies, ``merged stars'' or 
noise.  All objects classified as noise were excluded from further 
analysis.    

As a preliminary to identifying proper motion stars, all objects (stars and
galaxies) identified in the scans of the plates taken at the same epoch 
must be matched.  In the case of
the first epoch POSSI data, this is straightforward, since the 
O and E plates (corresponding roughly to photometric 
B and R observations) were taken on the same night.  
Thus there is no real motion between the time of the observations and the plate
scans can be matched by demanding positional coincidence to a 
high level of precision.    
For our second epoch TRM data, the IIIaF (photometric R) and IV-N 
(photometric I) plates were not observed on the same night.  
TRM paired the objects identified on their scans using a 3 arcsec box.  
Since the TRM plates were taken $\sim 3$ years apart, this procedure 
loses the highest ($\mu \gtrsim 1$ arcsec yr$^{-1}$) proper motion 
stars.  In our survey, this is not a concern because the 40 year 
separation between the first and second epoch, with no
plates at intermediate epochs, prevents our
identifying reliably stars with high proper motion 
due to the large number of possible pairings. The effective upper limit
for our survey is $\mu$=0.375 arcsec yr$^{-1}$, and that limit is taken
into account in our luminosity function analysis. 

We should note that the TRM detections are restricted to the 
region within 3 degrees of the plate center in order to avoid the
vignetted regions near the edge of the plates, and that the COSMOS
scans did not cover the entire plate.  The appropriate area
of the survey is given in Table~\ref{table-plates}. Moreover, the
the POSSI field centers do not correspond to 
the POSSII/UKST field centers.  Thus, between one and four POSSI plates
are paired with the TRM fields, as listed in Table~\ref{table-plates}.

\begin{figure}
\plotone{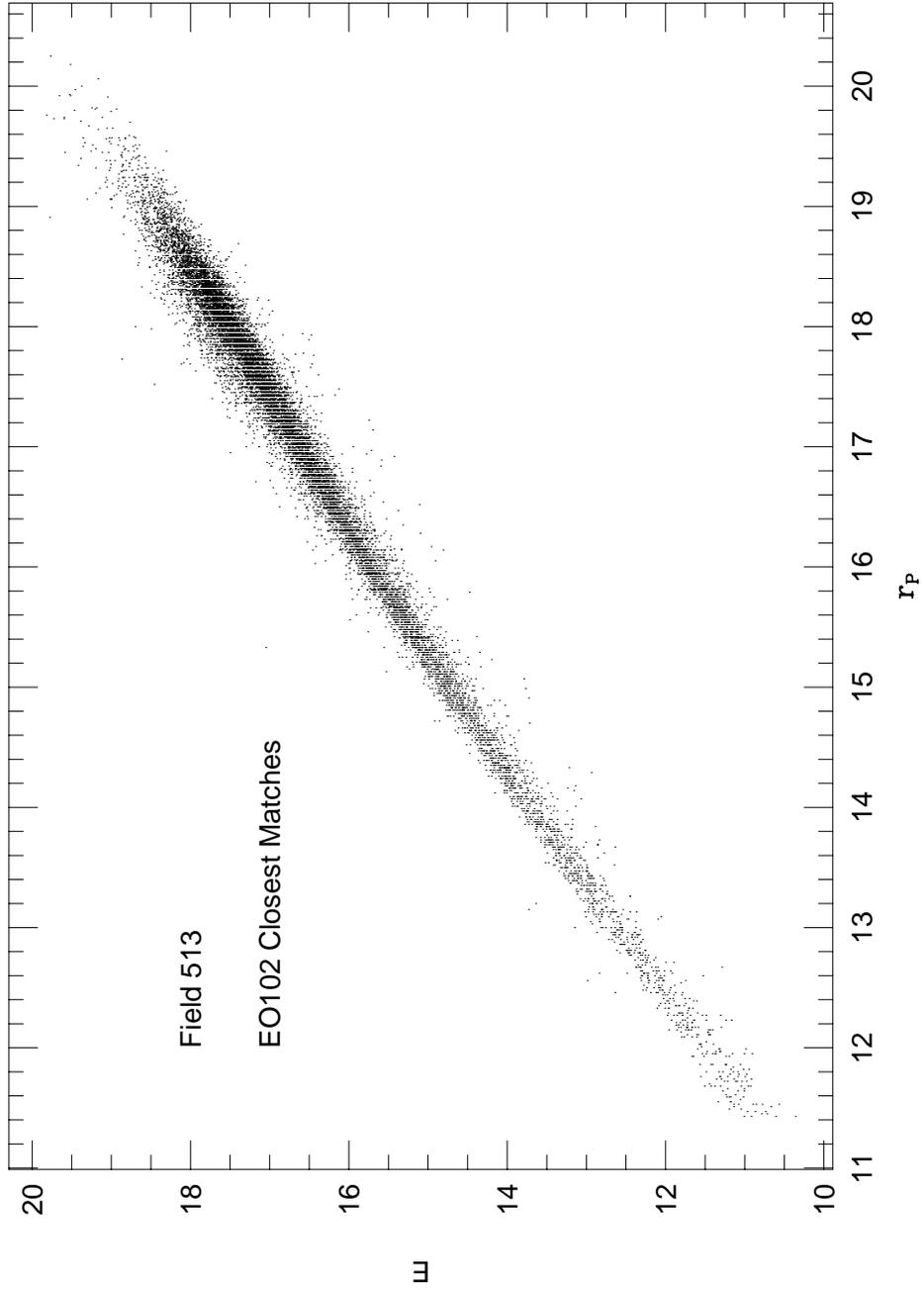}
\caption[Magnitude Comparison]
{Comparison of magnitudes for matches within 3 arcsec
for the POSSII field 513  and POSSI plate E102.  
The outlier points are due to galaxies and merged sources which are
treated differently by the two measuring machines.}
\label{figure-close513}
\end{figure}

The identification of the proper motion stars was made as follows.
First, an initial pairing of objects within a 3 arcsecond 
box ($\Delta <3"$) was made.
In each field, the surface density of celestial objects is sufficiently low
that matching within this radius results in an unambiguous pairing.
Figure~\ref{figure-close513} illustrates the agreement between the 
$r_P$ and E magnitudes for one representative pairing (POSSII Field 513 and
POSSI E102).  Most of the outliers evident in
this plot are correct pairings of extended objects 
(galaxies or merged stars) --- they are outliers due to 
differences in the definitions of magnitudes used by the two machines,
as well as the different plate characteristics.  
For this study, we are interested only in stars, so such 
outliers are unimportant.  
Having paired the unambiguous matches, we obtain a list of 
unmatched objects --- these objects have no counterparts within 3 arcseconds
at the second epoch and are good candidates for proper motion stars.  
Eliminating paired objects, we find all possible matches within 
15 arcsec of unmatched objects. We exclude pairings which have  magnitudes
inconsistent with the E --$r_P$ relation delineated by the 
close matches --- inconsistent being defined as a discrepancy 
of from $>$1 magnitude 
$r_P = 12$ to $>1.5$ magnitudes at
$r_P = 19$. These limits are ``conservative'' in the sense that some 
false pairings are included in order 
to avoid excluding true proper motion stars.

The vast majority of objects on the plate should show insignificant
proper motion between the two epochs, but directly matching the 
right ascension and declination catalog positions leads to 
many (false) proper motion objects. Those spurious motions arise from
systematic errors in the relative positions 
of the two catalogs (caused by differing plate centers, 
inadequacy of the plate model, plate variations, telescope differences, etc.)
These systematic positional errors can be corrected by determining the
transformation between the original pixel positions for the two epochs
($x_1,y_1$ and $x_2,y_2$).  Both our own experience with 
the TRM fields and the Reid (1990) analysis showed that a polynomial
solution over the entire plate works poorly, producing many false proper 
motion stars.  Analysis over smaller regions, however, is effective.  
Each plate was divided into approximately one by one degree regions
and the positional transformation determined 
using second-order two-dimensional polynomials.  Only stars 
with $\Delta < 3"$ matches were used for the initial 
co-ordinate transformation, iterating  twice to exclude objects with 
large ($>2 \arcsec$) residuals. We then used these polynomial solutions to 
determine the proper motions of {\emph all} stars in the field.
The stars that prove to have $\mu \ge 0.1$ arcsec yr$^{-1}$ are included 
in our sample of proper motion stars. 
The selection of the candidate
halo stars is described in Section~\ref{rpm}.  

\begin{figure}
\plotone{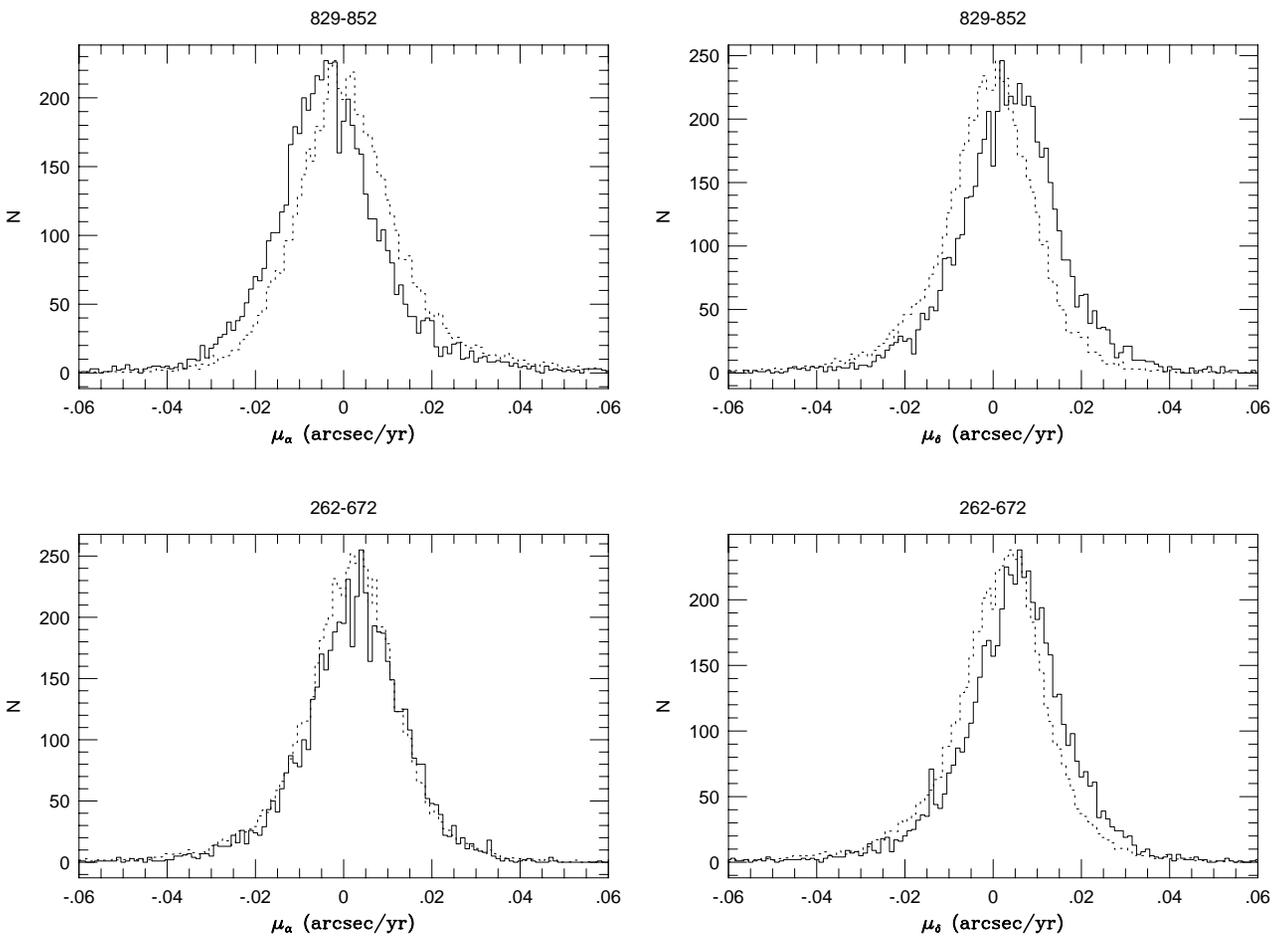}
\caption{Measured relative proper motions for galaxies and stars.
The amplitude of the stellar histograms  have been scaled down by a 
factor of $\sim 100$.  The median of
the galaxy distribution defined the correction to absolute proper motions.}
\label{figure-absmu}
\end{figure}

These proper motions are measured relative to faint
stars, which have typical distances of $\sim 500-4000$ pc.  
Thus we expect the reference frame to have a small, 
but non-zero, mean proper motion.
We derive the conversion from this relative frame to an absolute
(extragalactic) frame using measurements of objects classified as galaxies.
The latter are identified using the $\phi$-parameter
defined by Picard (1991).  We take $200<\phi<600$ for these purposes, which
provides a sample dominated by galaxies.  The upper cutoff is taken to 
exclude the ``fuzziest'' galaxies --- we found that these ``fuzzy''
galaxies have large spurious proper motions, which we attribute to a
combination of the different plate material used at the different epochs, 
and the different scanning and centroiding techniques 
used at the APM and COSMOS.
In Figure~\ref{figure-absmu}, we plot the distribution of galaxy proper
motions and stellar proper motions for the fields.  
It is clear that the galaxies
do indeed show a systematic shift with respect to the stars. 
Table~\ref{table-absmu} 
lists the median proper motions ($\Delta \mu$) derived for the galaxies, which 
at most amount to $5.2$ mas~yr$^{-1}$.  We have checked these values by
comparison with the projection of the solar motion and rotational
lag for the IPII and halo populations.  The direction of 
motion derived by the galaxies (i.e., the {\it sign} of the proper motions)
agrees with the expectations.  
These values given in Table~\ref{table-absmu} are used to correct
the relative proper motions to an absolute system:
$$\mu_{abs} = \mu_{rel} - \Delta\mu$$
We also give the observed standard deviation of the galaxian motions, with
values in the range $10.6 - 17.0$ mas~yr$^{-1}$.  
These provide an estimate of the uncertainties of our proper motions,
but we expect that the extended galaxies should have 
poorer positions, and hence less accurate proper motions, than is
the case for stars. 

It is evident in Table~\ref{table-absmu} that Field 831 shows the
largest random errors and has absolute proper motions
correction that are discrepant at the $\sim 2$ mas level from
the adjoining fields.  R. Mendez (private communication) has
kindly computed model proper motions for this direction based upon 
the galactic structure model of Mendez \& van Altena (1996).
The predicted mean stellar proper motions of 
($\mu_\alpha,\mu_\delta$) = ($3.8, -4.2$) mas~yr$^{-1}$ are in excellent
agreement with our observations in Fields 829 and 832.  We take this as confirmation
that our simple correction technique using galaxies is adequate for our
purposes.  The data for Field 831 are evidently of lower 
astrometric quality, but
the accuracy remains adequate for identifying the high proper
motion stars analyzed in this study. 

The NLTT catalog (\cite{nltt}) offers an opportunity to check both the 
completeness and accuracy of our measurements.
As noted by many authors, identifying the 
NLTT stars can be difficult since some of
the positions are quite poor (for example, we find a
star that matches the magnitudes and proper motion of 
LP 621-64, but is 10 {\em arcminutes} east of Luyten's position).  
Excluding stars which are too bright, too faint, or
too blue (i.e., too faint to be seen on the POSSII IV-N plates), 
we find that 122 of 137 NLTT stars are recovered, indicating 
a completeness of at least $89\%$.   
We find that the standard deviations of 
$\mu_{\alpha}^{POSS} -  \mu_{\alpha}^{NLTT}$ and  
$\mu_{\delta}^{POSS} -  \mu_{\delta}^{NLTT}$ are in the range
10 to 25 mas yr$^{-1}$.  Luyten estimated his motions to be
accurate to $\pm 15 - 25$ mas yr$^{-1}$.   The combination of
our uncertainties estimated from the galaxies and the NLTT uncertainties
account for the observed scatter.  

\subsection{Reduced Proper Motions\label{rpm}}

Selection of halo candidates using reduced proper motion  
requires photometry in addition to the proper motions.  
We have used the photometry from the second epoch plates which
provides us with R and I measurements for each star (except in
the NGP field, for which we have V and I).   

We adopt the photometric calibration determined by TRM using
CCD measurements for each plate.
Because their (``Palomar'') calibration is on a unique system, 
we describe it briefly here.  A complete description is given by TRM.
The best defined photometric system in R and I
for late-type stars is the Cousins system 
(\cite{cousins})\footnote{As in TRM, we use the ``R'' and ``I'' 
to denote generic R and I photometry on any system and indicate 
particular photometric systems with subscripts.  Thus R-I$_C$
is Cousins R-I, $(r-i)_P$ is TRM's Palomar system, and 
$(r-i)_G$ is Gunn R-I.}
However, M dwarfs have strong color terms in this system because
the R$_C$ band filter extends far into the I band --- as a result, the
effective wavelength of the R$_C$ filter moves redward for cooler
stars, as discussed by Bessell (1986).  The POSSII IIIaF emulsion
plus filter combination has a sharper long-wavelength 
cutoff at $\lambda 6900 \AA$, 
so the color term differs significantly from the 
standard Cousins system.  The POSSII response is close to the
Gunn r filter defined by Thuan \& Gunn (1976).  
Therefore, TRM's CCD calibrations of the fields used
Gunn r$_G$ and i$_G$ filters but observed {\it Cousins} late-type
standards.  These CCD magnitudes were matched to the
COSMOS photographic magnitudes.   
They found that no significant I color term was necessary
to calibrate the observations but, as expected, a strong R color term was 
necessary.  Thus, their photometric observations consist of
an R magnitude, denoted $r_P$, an I magnitude, denoted $i_P$,
and a color, denoted $(r-i)_P$, for each star.  
The relation between the standard Cousins color and the Palomar color is
\begin{equation}\label{ri-ri}
(r-i)_P = 0.162 + 0.439(R-I)_C + 0.671(R-I)_C^2 - 0.36(R-I)_C^3
+0.074(R-I)_C^4 
\end{equation}
\begin{equation}
i_P = I_C
\end{equation}
Note that TRM use the relation $i_P = I_C - 0.046$, but examination of
their Figure~2 indicates that there is no offset between $i_P$ and
$I_C$ for $(R-I)_C < 1.2$, which is the color of the halo stars
in our survey.  
The V and I photometric calibration for the NGP field is 
described in Reid (1990).

\begin{figure}
\plotone{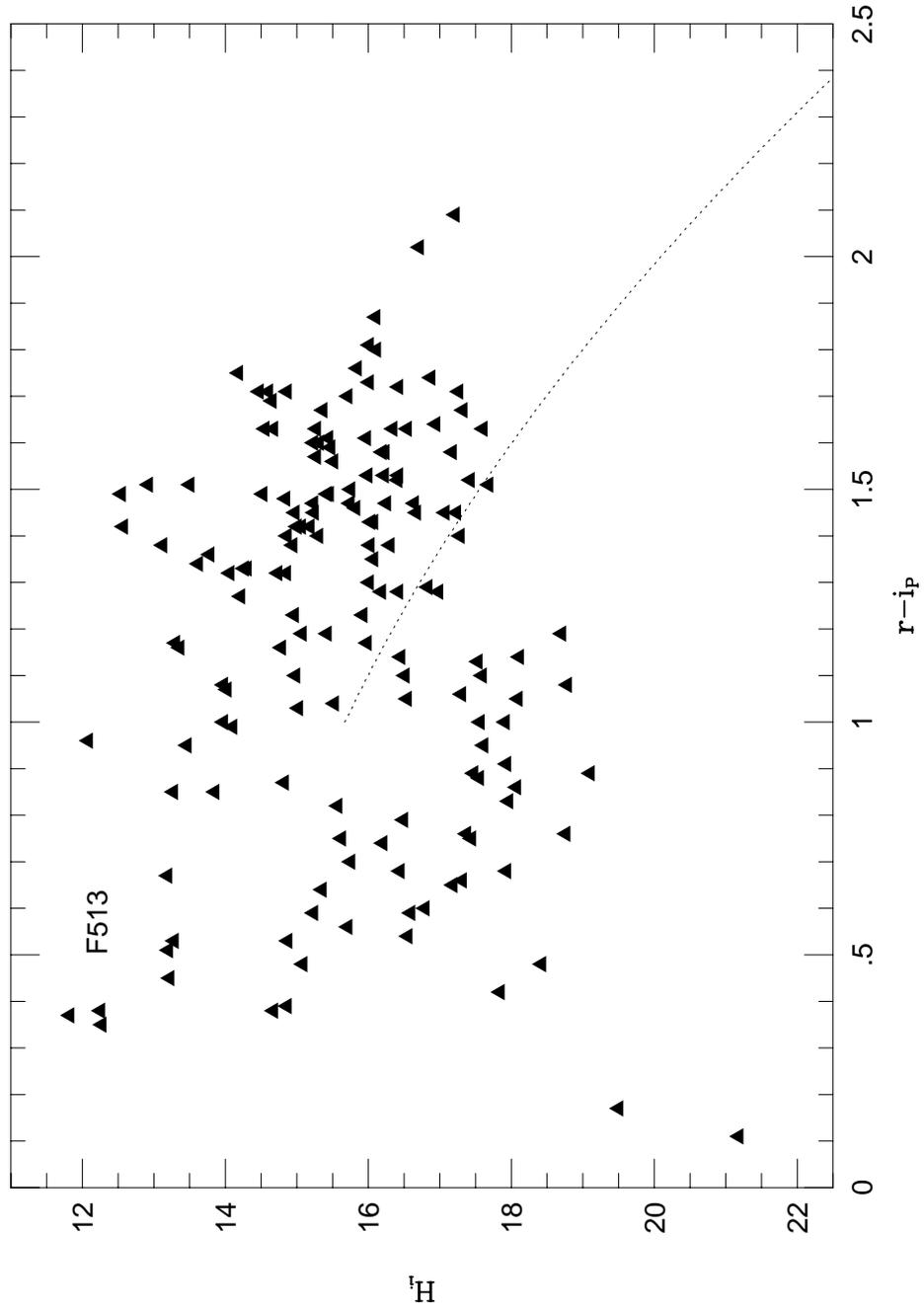}
\caption[RPM Diagram: Field 513]{The reduced proper motion diagram for 
Field 513.  Halo stars lie below the dotted line, which corresponds
to $220$ km s$^{-1}$ for near-solar metallicity disk stars.}
\label{figure-rpm513}
\end{figure}

The reduced proper motion diagram for Field 513 is shown in 
Figure~\ref{figure-rpm513}.  The solid line marks the 
disk main sequence defined by TRM, 
\begin{equation}\label{tinney-mainsequence}
M_{i} = 5.56037 + 0.458615 \times (r-i)_P + 1.32084 \times (r-i)_P^2
\end{equation}
offset to match $v_{tan} = 220$ km s$^{-1}$. Stars falling below this line
(larger H$_I$) are identified as
halo candidates for spectroscopic followup.  Recently, we have 
pointed out that the disk main sequence shows evidence for a 
``kink'' at spectral type M4.0 V (\cite{gr96}; \cite{rg97a}).
While this feature is not included in the TRM calibration, it
occurs at $(R-I)_C \approx 1.6$, so it is not relevant
to the current selection of candidate halo subdwarfs.

While our tangential velocity limit is set at $220$ km s$^{-1}$ for
solar abundance stars, M subdwarfs are selected 
at smaller $v_{tan}$.  The esdM enter the candidate pool 
for $v_{tan} > 75$ km s$^{-1}$ and the brighter sdM at
$v_{tan} > 125$ km s$^{-1}$.  The most likely contaminants,
modestly metal-poor ($[m/H] \approx -0.6$) 
Intermediate Population II M dwarfs, lie 
$\sim 0.5$ magnitudes below the disk main sequence
(\cite{g97}, Figure 7), and therefore enter at $v_{tan} > 175$ 
km s$^{-1}$.

\subsection{Spectroscopy\label{spectroscopy:halolf}}

We obtained optical spectra of the candidate halo stars in order to
determine radial velocities and metallicities.  
The Hale 200-in. and the Las Campanas Du Pont 100-in. telescopes
were used to obtain data in the wavelength range 
$6100 - 7300 \AA$ with $\sim 3 \AA$ resolution.  
These parameters were chosen to correspond to the 
observations of nearby M dwarfs (\cite{rhg95}; \cite{hgr96})
and M subdwarfs (\cite{g97}) that we have already published.    
For the 200-in. observations, we used the Double Spectrograph
(Oke and Gunn 1982).  
In August 1995,  the blue camera was set to 
observe $6000-6900 \AA$ and the red camera was set to  $6700-8000 \AA$
using $600$ l/mm gratings blazed at $4000 \AA$ and $10000 \AA$ respectively. 
In October 1995, a new red camera was installed in the
double spectrograph and was used in all subsequent runs to observe the region
$\lambda 6000-7400 \AA$ at $1.4~\AA~{\rm pix}^{-1}$ with the
 $600$ l/mm grating blazed at $10000 \AA$.  With this setup,
the blue camera was used to cover $4000 - 5500 \AA$ with a
$300$ l/mm grating blazed at $3990 \AA$.  In practice, the latter
data were useful only for the bluest ($(r-i)_P < 0.6$) stars.  
The Du Pont observations used the modular spectrograph 
with a 1200 line grating  blazed at $7500 \AA$.  

From these spectra, we determined radial velocities and spectral types.  
During each observing run, we observed K and M radial velocity standard stars 
drawn from the list of Marcy \& Benitz (1989).  A velocity for each
candidate halo star was determined by cross-correlating with the 
best match standard star.  Comparison of subdwarfs with known
radial velocities observed during the same runs shows that the accuracy is 
$\pm 20$ km s$^{-1}$ (\cite{g97}).  Using these radial velocities,  we
measured TiO and CaH bandstrength indices, as defined originally
by Reid {et al.} (1995).  Each index measures the ratio
of the flux ($F_{\nu}$) within the absorption feature to nearby 
pseudo-continuum points.  Table~\ref{table-index} lists the 
wavelength definitions of the features (W) and the 
pseudo-continuum (S1,S2) points.  We have defined a spectral 
classification system using these indices
(\cite{g97}).  Stars are classified
as disk stars (M V), M subdwarfs (sdM), or extreme M subdwarfs (esdM).  
We have shown on the basis of model atmosphere calculations (\cite{ah95}) 
that these empirical classifications correspond to 
$[m/H] \approx -0.5$, $[m/H] \approx -1.2\pm 0.3$, 
and $[m/H] \approx -2.0 \pm 0.5$ (where $[m/H]\approx[Fe/H]$;
\cite{gr97}).  Each halo candidate  has been classified using this system.  

\section{The Luminosity Function\label{lf:halolf}}

In this section, we calculate the I-band luminosity function 
from the data described in Section~\ref{data:halolf}.  
The luminosity function, $\Phi(M_I)$, is defined as the
number of stars per cubic parsec per magnitude in a bin 
centered at $M_I$ .  We use Schmidt's (1968) $1/V_{max}$ technique to estimate $\Phi(M_I)$
\begin{equation}\label{equation-firstlf}
\Phi_{obs} = \sum \frac{1}{V_{max}}
\end{equation}
\begin{equation}
V_{max} = \frac{\Omega}{3}(d_{max}^3 - d_{min}^3)
\end{equation}
Where $\Omega$ is the solid angle on the sky of the survey, 
$d_{max}$ is the maximum distance that a given star could have
been detected in this survey, and $V_{max}$ is the corresponding
volume.   A minimum distance, $D_{min}$, also appears to account for
the distance at which the star would have evaded detection due to the
upper proper motion limit of $\mu_{max} = 0.375 \arcsec$ yr$^{-1}$.  
The maximum distance may be limited by either
the lower proper motion limit ($\mu_{min} = 0.100 \arcsec$ yr$^{-1}$) or 
the limiting magnitudes ($R_{0},I_0$), hence for a star, distance d, 
magnitudes (R, I), 
\begin{equation}\label{equation-lastlf}
d_{max} = d \times \rm{Min}\left({{\mu}\over{\mu_{min}}}; 
\rm{dex}[0.2(R_0-R)]; \rm{dex}[0.2(I_0-I)]\right)
\end{equation}
We estimate the associated error by assuming that each star contributes
an uncertainty of $1/V_{max}$ (\cite{felton}).  Then
\begin{equation}   
\sigma_{\Phi}^2 = \sum \frac{1}{V_{max}^2}
\end{equation}

As noted in Section~\ref{general:halolf}, we apply a tangential 
velocity cutoff in order to ensure a ``pure'' halo sample. We
adopt a value of $v_{cutoff}=200$ km s$^{-1}$ in our main analysis. 
As a result, 
a correction must be applied to the apparent luminosity function
in order to obtain the space of density of halo stars.  
Thus,
\begin{equation}
\Phi_{halo} = \frac{1}{\chi}\Phi_{obs} 
\end{equation}
The discovery fraction $\chi$ is a function of position on the sky
and the adopted $v_{cutoff}$.
In our case, we compute $\chi$ for each field in the survey. 
This procedure is described in Section~\ref{kinematic:halolf}.

A check on the completeness of the survey can be made by considering
\begin{equation}
\left\langle \frac{V}{V_{max}} \right\rangle = 
\left\langle \left(\frac{d}{d_{max}}\right)^3 \right\rangle
\end{equation}
which considers the ratio of the volume $V$
corresponding to the star's actual distance $d$ to $V_{max}$.
In the case of a spatially uniform sample, such as the halo over
the volume probed by this survey, we expect 
$\left\langle \frac{V}{V_{max}} \right\rangle = 0.5$.  
We find that eight of our nine fields are within $1 \sigma$ of
0.5 for sdM/esdM with $v_{tan} > 200$ km s$^{-1}$, and 
conclude that, as a whole, the survey is complete statistically.

\subsection{Adopted Color-Magnitude Relations\label{section:colormag}}

\begin{figure}
\plotone{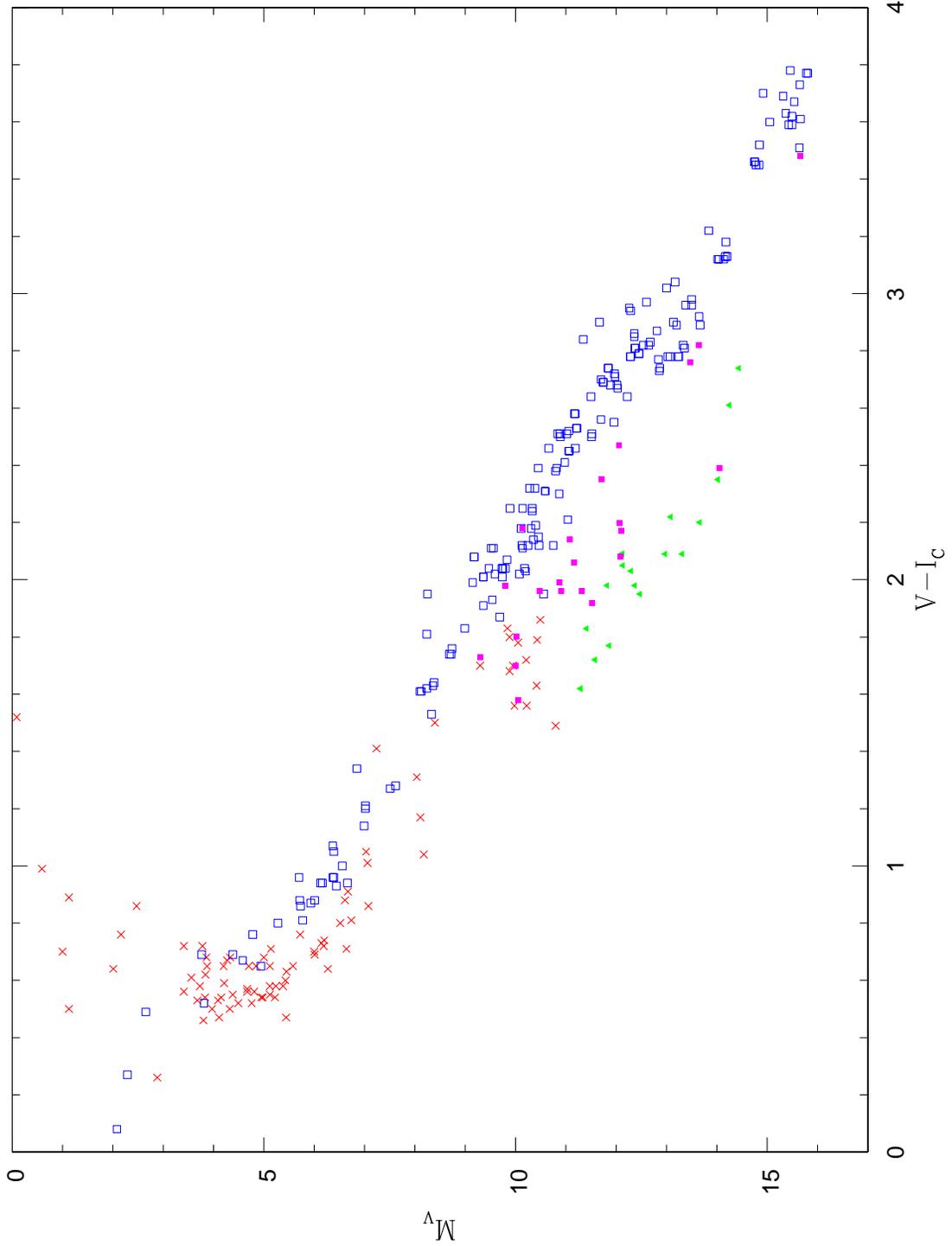}
\caption[The Population I and II Main Sequences]{The
HR Diagram for stars using trigonometric parallaxes.
Crosses are stars earlier than type M with $v_{tan} > 220$  km s$^{-1}$ from 
the Yale Parallax Catalog (\cite{ypc4}) or the Hipparcos Catalog,
solid squares are sdM (Chapter 2), and solid triangles are esdM.  
The disk sequence is shown using nearby stars (open squares)}  
\label{figure-hipp}
\end{figure}

We use the photometry and parallax data compiled by Gizis (1997)
to determine color-magnitude relations appropriate for 
each spectral class of star (M V; sdM; esdM).  There are many 
sdM and esdM with accurate distance determinations 
available, mainly due to the efforts of 
the USNO CCD parallax program (\cite{m92}), which allows us to
determine the following relations:
\begin{equation}\label{ms-vi-sdm}
M_I^{sdM} = 4.24 + 2.40 \times (V-I)
\end{equation}
\begin{equation}\label{ms-vi-esdm}
M_I^{esdM} = 4.58 + 2.92 \times (V-I)
\end{equation}
These relations are valid for $(V-I) \ge 1.6$.  Our survey is 
aimed at the fainter, redder M subdwarfs, but for comparisons to
other surveys of higher mass stars 
it is useful to have an estimate of the main sequence for the bluer
subdwarfs.  For these stars, the above linear relations are
no longer appropriate, due to an inflection in the main sequence at
$V-I \sim 1.5$ (c.f., \cite{dm96}; \cite{bcah97}).  Unfortunately, there are
few parallax K subdwarfs and fewer with red (RI) photometry.
The Hipparcos mission, however, has determined parallaxes for 
many G subdwarfs.  We therefore fit a linear relation to the Hipparcos G
subdwarfs supplemented by the available K subdwarfs from the
Fourth Edition of the Yale Parallax Catalog (van Altena {et al.} 1996).
In this region, models predict that for $[m/H] \lesssim -1$ the
absolute magnitudes are relatively insensitive to metallicity (\cite{bcah97}). 
We select only stars with $v_{tan} > 220$ km s$^{-1}$  to ensure a halo sample.
The data are shown are in Figure~\ref{figure-hipp}.  There are 29 subdwarfs
with ${{\sigma_\pi}\over{\pi_t}} < 0.2$, $5 \le M_V < 10$, and 
$0.5 < V-I < 1.4$.  We find
\begin{equation}\label{ms-vi-sdg}
M_I^{esdG/sdG} = 2.99 + 3.16 \times (V-I)
\end{equation}
This slope is consistent with the slope measured in globular clusters
by Santiago {et al.} (1996).  

The above relations can only be used for the NGP field, where
we have V and I photometry.  Unfortunately most sdM and esdM lack
R photometry, while we have only R and I data for the remaining fields in
our survey.  Hence we have determined the relationship between
$(V-I)_C$ and $(r-i)_P$ using Equation~\ref{ri-ri} 
and Bessell's (1990) VRI$_C$ photometry of nearby stars.  We find:
\begin{eqnarray}
(r-i)_P & = & 0.1477 + 0.4520 (V-I)_C - 0.3932 (V-I)_C^2+ \nonumber \\
& & 0.3606 (V-I)_C^3 -0.1020 (V-I)_C^4 + 0.009343 (V-I)_C^5
\end{eqnarray}
Applying this transformation to the $(V-I)_C$ color of each parallax star,
we find that 
\begin{equation}\label{ms-ri-sdm}
M_I^{sdM} = 5.37 + 3.53 \times (r-i)_P
\end{equation}
\begin{equation}\label{ms-ri-esdm}
M_I^{esdM} = 5.96 + 4.29 \times (r-i)_P
\end{equation}
Relations~\ref{ms-ri-sdm} and ~\ref{ms-ri-esdm} are valid 
for $(r-i)_P \ge 0.77$.  It is possible that the R-I,V-I 
transformation for the extreme M subdwarfs is not the same as 
that for disk M dwarfs. While the 
currently available photometry (summarized in \cite{g97}) is unable to
resolve this issue, we have obtained spectrophotometry 
(\cite{mythesis}) which indicates that there is no significant color
term for the sdM stars or the earliest (esdK7; esdM0.0) extreme subdwarfs.  

For the G subdwarfs, we transform
Equation~\ref{ms-vi-sdg} and
obtain:
\begin{equation}\label{ms-ri-sdg}
M_I^{esd/sdG} = 1.98 + 8.26 \times (r-i)_P
\end{equation}
Equation~\ref{ms-ri-sdg} applies for $0.31 < (r-i)_P < 0.66$.

\subsection{Kinematic Corrections\label{kinematic:halolf}}

The luminosity function computed using the techniques described
above, although correct in shape, represents only the fraction of 
the halo that has $v_{tan} \ge v_{cutoff}$ km~s$^{-1}$.
Given a kinematic model of the halo, we can calculate 
the discovery fraction ($\chi$) for each field using Monte
Carlo simulations.  BC have shown that variations of
$\pm 5$ km~s$^{-1}$ in the kinematic parameters 
($\langle V \rangle$,$\sigma_U$,
$\sigma_V$,$\sigma_W$) imply variations of less than $4\%$ in $\chi$;
sadly, published data for halo stars span a much larger range of
kinematics.  We therefore compare the observed kinematics of
our survey stars to the predictions of representative halo models. 

Analyses of local (d$<$ 1 kpc) halo stars generally derive
mild prograde rotation (e.g., $37 \pm 10$ km s$^{-1}$, 
Norris 1986), although recent studies based on more distant ($>$5 kpc) 
halo stars have found evidence for a retrograde 
rotating halo (see the review of Majewski 1993).
We compare our data to the models of Bahcall \& Casertano (1986, BC), 
Norris (1986, N), Casertano {et al.} (1990, CRB), Layden {et al.} (1996, L),
and Beers \& Sommer-Larson (1995, BSL).  The last is of considerable
interest as it features a retrograde rotating halo.  

For each of these models, we generated a Monte Carlo 
simulation of our survey with $\sim 5000$ detected stars.
The halo velocities are represented by Gaussian distributions 
in U,V, and W.\footnote{We use the standard notation of (U,V,W)
for the space velocity components, in which U represents the
motion towards the Galactic center ($l=0$,$b=0$), V represents
motion in the direction of Galactic rotation, and W represents
motion perpendicular to the Galactic plane.}  
In Table~\ref{table-compareuvw}, we compare the predictions
of the U,V, and W velocity distributions of the models 
to the actual data for the 35 metal-poor stars with $M_I > 8.5$.  
We emphasize that 
these are the predicted velocity characteristics of the 
kinematically-biased observed sample, not the true halo kinematics.
We also give the probability that the observed data and the model simulations
are drawn from the same distributions, as determined by 
Kolmogorov-Smirnov tests.  The comparison is most sensitive to the
observed V velocities, which are most important in determining
$\chi$.  It clear that all the published models
agree fairly well with the observed U and W velocity distribution,
although the N model is significantly poorer for U, due to
its smaller velocity dispersion ($\sigma_U = 131$ km s$^{-1}$).
The V velocities favor prograde halos, with the 
BC model ($v_{rot} = 66$ km s$^{-1}$) yielding too small a mean
V velocity and the N model ($v_{rot} = 37$ km s${-1}$) yielding too 
large a mean velocity.  The BSL model's retrograde halo is much less
likely than the prograde models. Of the published models,
the Layden {et al.} model gives the highest probability when all three
velocity components are considered.  

Guided by the comparison above, we have constructed a composite
kinematic model, taking the parameters derived by Norris as a starting point
and increasing both $\sigma_U$ to match other studies, and the halo rotation
to a value midway between BC and N.  The resulting parameters are
$(v_{rot},\sigma_U,\sigma_V,\sigma_W) = (+50, 140, 106, 85)$ in
km s$^{-1}$ and we refer to this as our model A.
These changes are within $1.5 \sigma$ of Norris's
estimated uncertainties. The values of $1/\chi$ for the BC and N models
lie within $\pm 7\%$ of the composite values, which are listed in 
Table~\ref{table-corr}. Values of $1/\chi$ based on the 
Layden {et al.} halo model are on average $20\%$ smaller.  

We have also calculated models including the IPII (BSL, CRB).
Only one in $\sim 400$ IPII stars will enter our
$v_{tan} > 220$ km s$^{-1}$ sample 
(although we caution that this number depends upon the assumption of Gaussian
kinematics).  Since the IPII consists of approximately 
$2-5\%$ of stars locally, it will contribute less than $10\%$
of our observed halo sample and thus will not bias significantly
our measurements.  
This can also been seen in Table 1 of BC, where it is shown that
the kinematic properties of stars in Eggen's (1979a, 1980) 
proper motion catalog 
with $v_{tan}\ge 200$, $v_{tan}\ge220$ and $v_{tan} \ge 250$ km s$^{-1}$
are indistinguishable.  We expect that the IPII contribution is
much less than this, since direct spectroscopic observation shows that 
most IPII M dwarfs both locally
(nearby stars with $100 \le v_{tan} \le 220$ km s$^{-1}$, Gizis 1997)  
and at height of a few kiloparsecs (\cite{keckcounts}) 
are not classified as sdM or esdM.  We reject all spectroscopic M~V stars from
our sample, and therefore we expect negligible IPII contamination. 

For each field, we calculate the
luminosity function using the Equations~\ref{equation-firstlf}
to~\ref{equation-lastlf} using only stars with $v_{tan} \ge
200$ km s$^{-1}$.  
We combine data for all fields to derive our best estimate luminosity
function, calculated for 0.5 magnitude bins, which is listed in 
Table~\ref{table-lf}.  

These conservative selection criteria ensure a pure Population II halo
sample but suffers from the uncertainty of the kinematic correction.
Since we have obtained spectroscopy for all of our
reduced proper motion selected stars, we have information on
the actual space density of metal-poor very-low-mass stars 
with lower velocities ($v_{tan} > 100$ km s$^{-1}$ for esdM;
$v_{tan} > 125$ km s$^{-1}$ for sdM).  We have therefore computed
the space density of sdM and esdM separately, using the appropriate 
velocity cutoff for each metallicity class. The completeness corrections,
$1 \over \chi$, for these velocity cutoffs are only $\sim 10-20 \%$, 
since virtually all halo stars have $v_{tan} > 125$ km/s. 
The resulting luminosity functions for the sdM and esdM are listed in 
Table~\ref{table-lf-lowvtan}. This luminosity function 
is remarkably consistent with that estimated from the 
corrected ``pure'' halo sample.  The normalizations differ
by no more that $20\%$ from that predicted by our
composite model.  We discuss these low-$v_{tan}$ samples in more detail 
in Section~\ref{discussion:halolf}.  
 
\section{Discussion\label{discussion:halolf}}

\begin{figure}
\plotone{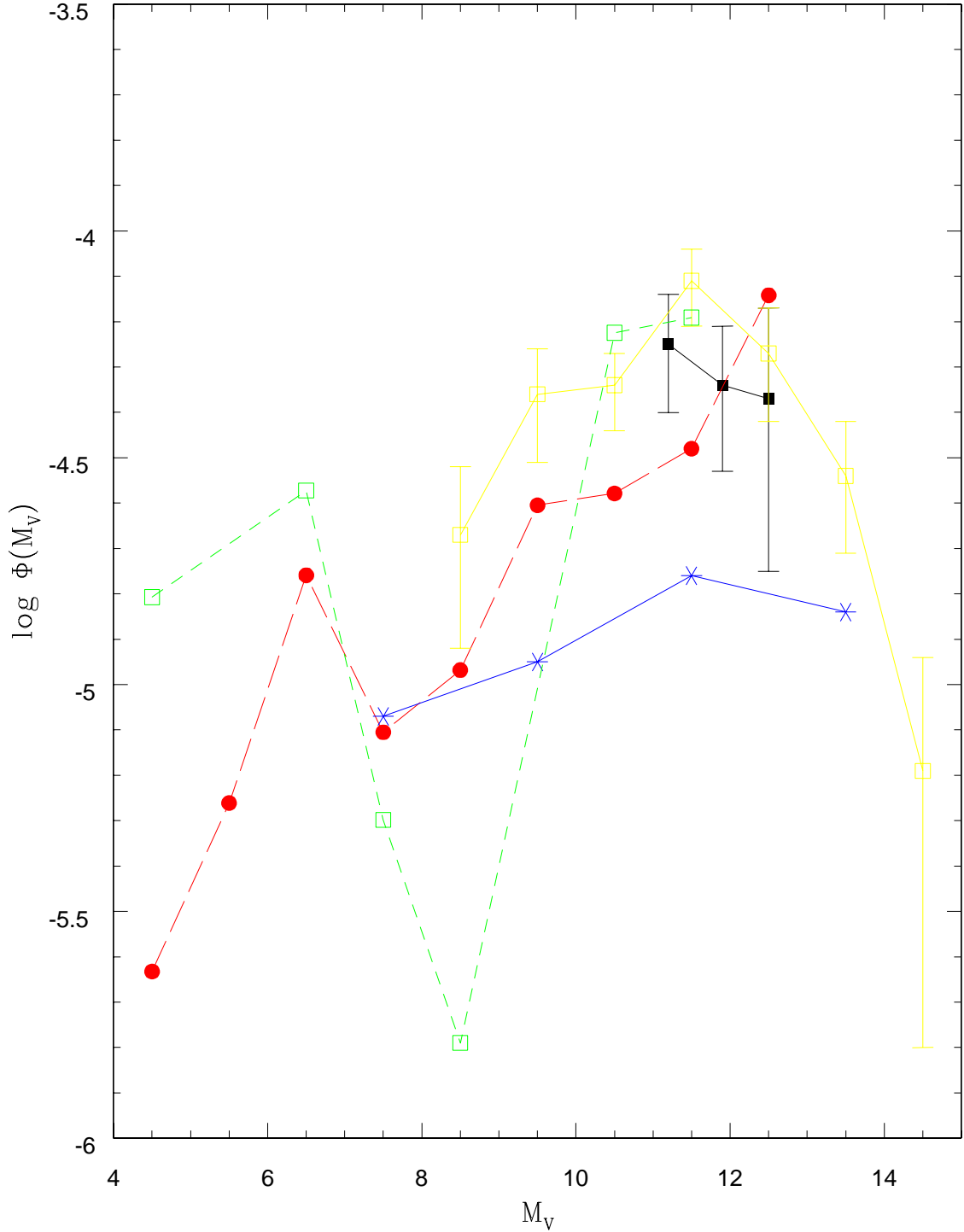}
\caption[$M_V$ Luminosity Functions]
{\small Comparison of our halo luminosity function with 
previously published halo V-band luminosity functions.  
Our luminosity functions, measured in $M_I$ have been transformed
to $\Phi(M_V)$ using Equations~\ref{ms-vi-esdm} (esdM) and
~\ref{ms-vi-sdm} (sdM) and then added together (solid squares).  
The luminosity functions shown are 
Schmidt (1975, open squares with dashed line), 
Bahcall \& Casertano (1986, solid circles with long-dashed line), 
Dahn {et al.} (1995, open squares with solid line),
and Gould {et al.} (1998, asterisks).  Note that our luminosity function
supports the higher value measured for local LHS stars by Dahn {et al.}
rather than the lower value found by Gould {et al.}.  
Also, even using a correction factor 3.0 for BC instead of
$\sim 2.5$ as in the other studies, BC's luminosity function apparently
underestimates the local space density of halo stars.}
\label{figure-field-mv}
\end{figure}

Our measurement of the Population II luminosity function
is aimed at only the limited range in luminosity corresponding
to the fainter M subdwarfs. Nonetheless, these results offer a means
of resolving the difference, noted in the Section~\ref{halolf-intro},
between Gould et al.'s HST starcount analysis  at $\gtrsim 5$ kpc 
and Dahn et al.'s local trigonometric parallax sample. Before discussing 
that issue, we consider whether our results can be combined with previous measurements
of the bright end of the halo luminosity function.  Since previous
studies of the halo luminosity function have presented
$\Phi(M_V)$ for all metallicities combined, we have
transformed our data using the relation
\begin{equation}
\Phi(M_V) = \Phi(M_I) \frac{dM_I}{dM_V}
\end{equation}
and the color-magnitude relations given in Section~\ref{section:colormag}.  
The results for the combined sdM and esdM luminosity function
is plotted in Figure~\ref{figure-field-mv}.  The other luminosity
functions shown are the local trigonometric parallax samples of Schmidt (1975)
and Dahn et al., the local photometric parallax sample of
Bahcall \& Casertano (1986), and the Gould et al. 
HST starcount analysis.  
We have not plotted Dawson's (1986) statistical
analysis of the LHS stars, since it is superceded by 
Dahn et al.'s calculations, which have the advantage of improved data 
for individual stars. Schmidt's results
have been updated using literature V-band photometry, but there are
only 0-2 stars per bin, except for the last two bins which have 
five stars apiece.  It should also be noted that, except for
the HST data, these estimates of $\Phi$(M$_V$) are based on
kinematically-selected samples. Dahn et al's $v_{tan}$
correction factors, $\chi$, agree to within 5\% with the values
predicted by our model A, and we have also used that model to 
adjust Schmidt's data. The Bahcall \& Casertano $\Phi$(M$_V$) plotted
is derived using their published correction factors. 
Adopting $\chi$ appropriate to model A reduces the inferred space densities by
$15\%$, increasing the discrepancy with respect to the other
studies (including our own results).

It is obvious from Figure~\ref{figure-field-mv} that there are 
potential problems involved in combining data from the several 
investigations to derive a halo luminosity function spanning the
full range in absolute magnitude. Of the surveys which include
brighter stars near the turnoff, Schmidt's luminosity function
(as noted originally by Schmidt!) has too few stars
to measure reliably the shape of the luminosity function.
The BC dataset has sufficient stars, but there is a clear,
systematic offset of almost a factor of two between their derived
space densities and Dahn et al's results. Bahcall \& Casertano's analysis
is based on Eggen's (1979a, 1980) survey of southern ($\delta < 30$) stars
with $\mu > 0".7$ yr$^{-1}$ and V$< 15$, but a comparison of both the 
relative numbers and $V/V_{max}$ distributions of such stars shows no evidence
for any significant increase in incompleteness at southern declinations. On
the other hand, the color-magnitude relation (from Eggen, 1979b) which BC 
use to estimate distances is a poor description of the lower Pop. II
main sequence (M$_V \gtrsim 9-10$), and therefore that portion of $\Phi$(M$_V$)
should probably be excluded. The calibration is likely to be more
reliable for the more luminous stars, since the main sequence is consistent
with the Hipparcos data in Figure~\ref{figure-hipp} for
the G subdwarfs.  

One notable aspect of the Eggen/BC sample is
that it may be incomplete in the Galactic Plane, since 
it is difficult to identify proper motion stars against
the dense stellar background.  We have simulated
halo samples using the BC, A, and L models 
for both ``all-sky'' ($\delta < 30$) and ``non-plane''
($\delta <30$ and $|b| >20$) selection. The kinematic correction factors 
($1/\chi$ for $v_{tan} > 220$ km s$^{-1}$) for the ``non-plane'' samples 
are greater than the ``all-sky'' samples 
by between 8\% and 12\% depending on the model.  
However, our analysis of the Eggen (1979a, 1980) data
shows that the derived halo space density for stars with
$|b| > 20$ is actually $\sim 40\%$ higher than 
that derived for the entire Eggen sample.  This suggests
that the BC luminosity function is underestimated by 
$\sim 30\%$ due to incompleteness in the Galactic Plane.  
This would account for about a third of the discrepancy
evident in Figure~\ref{figure-field-mv}.  

\begin{figure}
\plotone{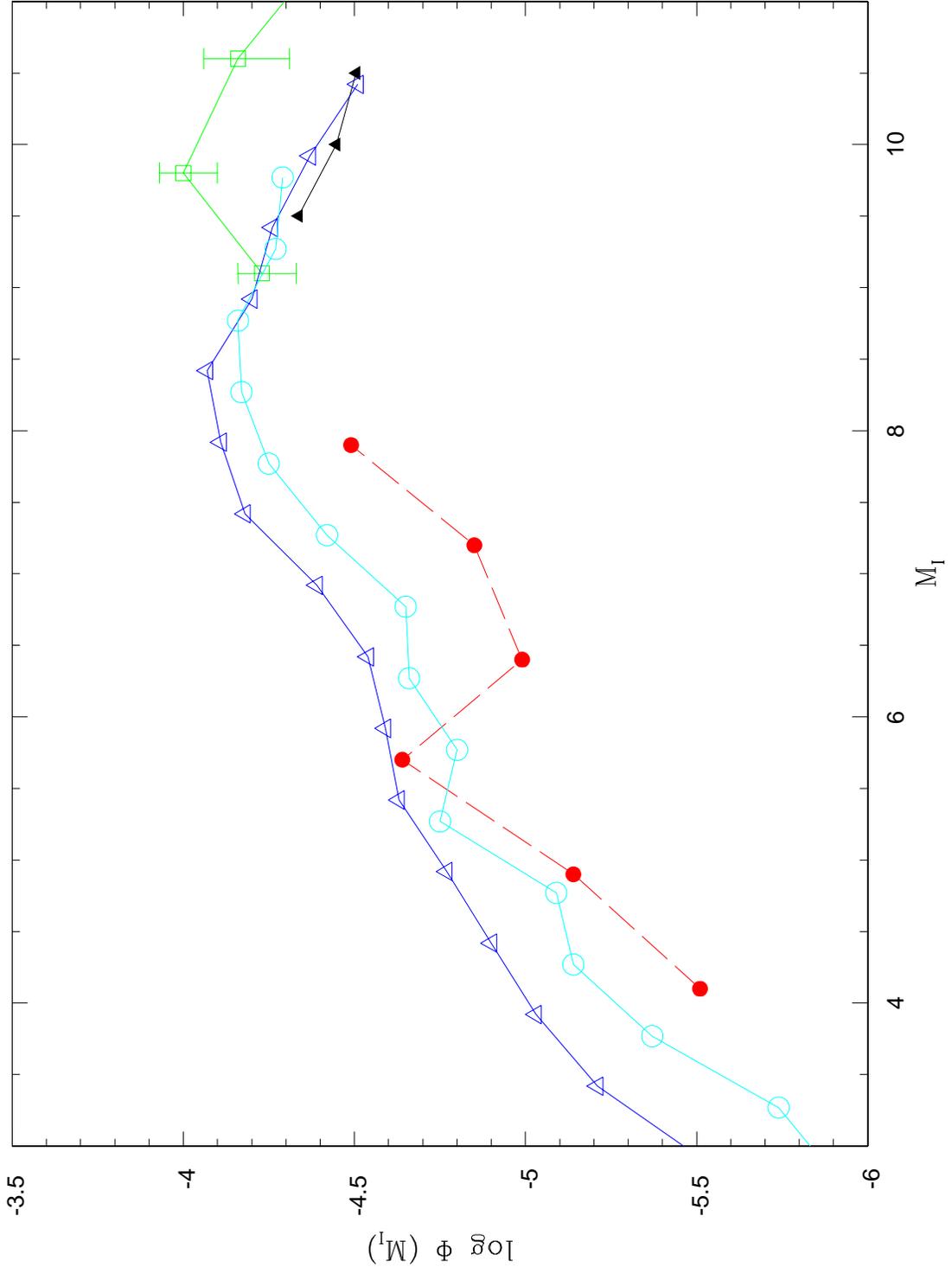}
\caption{\small The globular clusters NGC 6341 (open triangles) and 
NGC 7099 (open circles) compared to the esdM luminosity function
(solid triangles), Dahn et al. (open squares), and
Bahcall \& Casertano (solid circles).   The BC 
luminosity functions has been transformed to $\Phi(M_I)$, no account
has been taken of any metallicity spread; BC's original
kinematic correction has been applied.  Correcting for the
latter two effects would shift the BC luminosity function
downwards.  When the globular clusters are normalized
to our luminosity function, their luminosity is parallel to
but above the BC luminosity function for brighter stars.  
This is consistent with the apparent offset of BC compared to
to Dahn et al.  The Dahn et al. luminosity function for $M_I>9$ only
has been transformed using Equation 12.}
\label{figure-gci}
\end{figure}

Naively combining the higher-luminosity stars from BC with Dahn et al. and our
own results for fainter stars suggests a rather steep luminosity function.
Globular clusters provide a potential means of testing this result.
Figure~\ref{figure-gci} compares $\Phi$(M$_I$) from the field against
results for two of the better-studied clusters, using data 
from Piotto et al. (1997) adjusted to be consistent with the
new distances derived by Reid (1997) on the basis of main-sequence
fitting.  That comparison
suggests that the BC results may underestimate the space densities
of stars with $6.5 < M_I < 8$, consistent with the offset between
the BC and Dahn et al results plotted in Figure~\ref{figure-field-mv}.
It should of course be noted that the
field and cluster luminosity functions may differ -- the present-day
luminosity (mass) function of globular clusters could be 
affected by dynamical evolution, or alternatively the
initial mass function of field and cluster stars could
be different.  Extension of
the Dahn et al. sample to brighter absolute magnitudes 
(e.g., expanding the Schmidt sample) should 
provide the most effective means of confirming the shape of the field
luminosity function.    The Dahn et al. luminosity function, transformed
via Equation~\ref{ms-vi-esdm} is also plotted, but the transformation
is uncertain since we have assumed all stars to be esdM, which is
not the case --- and in any case, this preliminary 
Dahn et al. luminosity function represents a range of metallicities, 
so it may not be fair to compare it to the
globular clusters.  Nevertheless it is clear that the Dahn et al.
luminosity function peaks at a fainter absolute magnitude than 
the globular clusters.  

\begin{figure}
\plotone{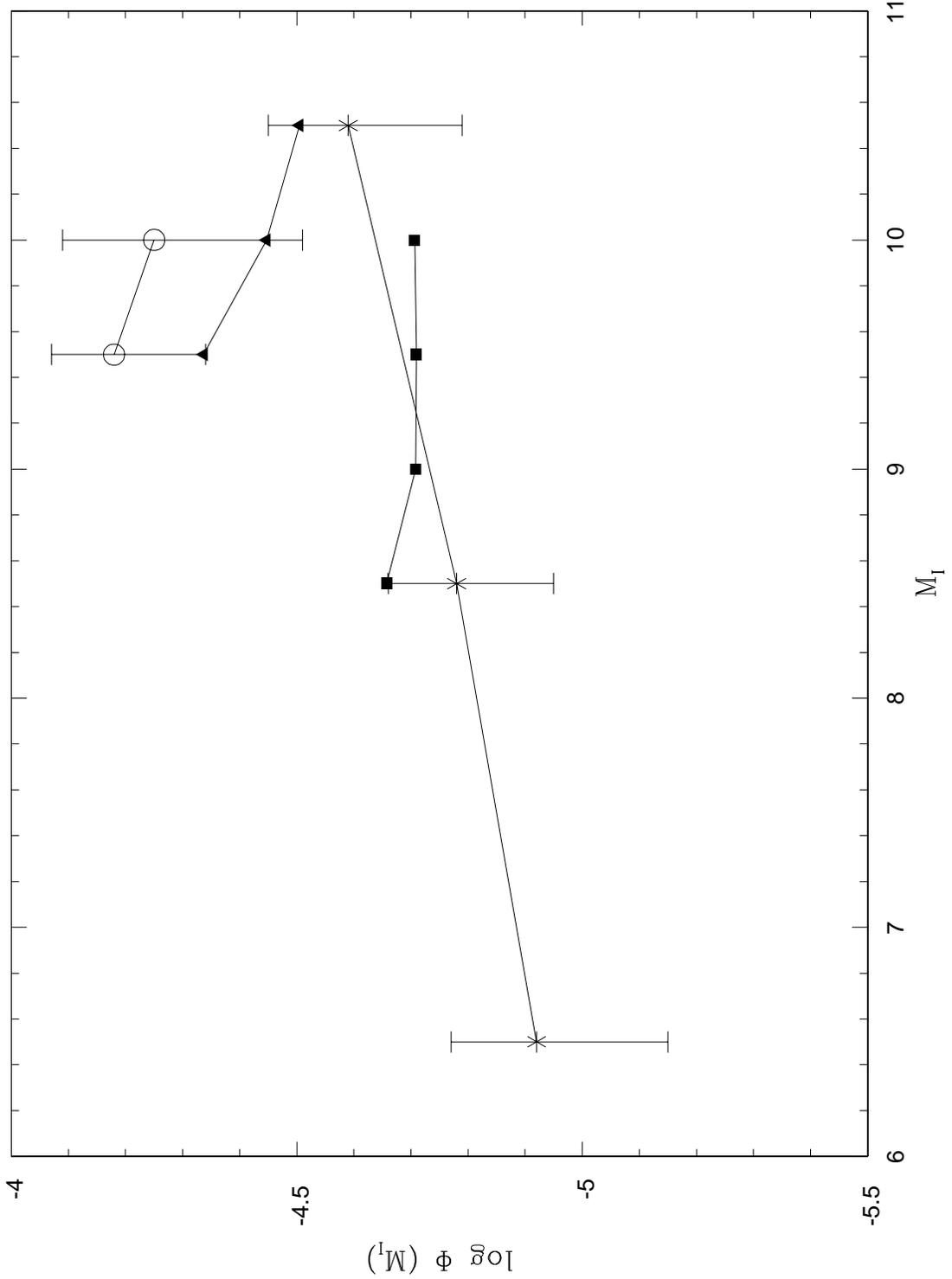}
\caption[I luminosity Functions]{Our estimated luminosity function for
sdM (solid triangles), esdM (solid squares), and both
combined (open squares).  Also shown is the Gould et al. HST luminosity
function (asterisks).  Due to crowding, the error bars have been suppressed
for the individual esdM and sdM datasets.  It is clear that the HST
luminosity function predicts substantially fewer stars than 
are actually seen. }
\label{figure-field-mi}
\end{figure}

As noted above, Figure~\ref{figure-field-mv} suggests that the
Gould et al (1998) HST analysis underestimates the local
density of late-type halo subdwarfs by a factor of four.
To underscore that point, Figure~\ref{figure-field-mi} plots our luminosity 
function against a  $\Phi(M_I)$ representation of the HST data that was 
computed by Gould (private comm.)
using the same data and techniques used to derive the $M_V$ luminosity
function of Gould {et al.} (1998).  While our study
is limited to a restricted range of luminosity, the space densities,
derived from a completely independent sample of nearby subdwarfs, are 
clearly strongly supportive of the LHS-star analysis. 
A possible resolution of the discrepancy between the distant-halo
and local-halo analyses may lie in the composite nature of the
galactic halo. Sommer-Larsen \& Zhen (1990) estimate that at least 40 percent
of the local subdwarfs reside in a highly-flattened component, which
would essentially be absent from the HST dataset. Taken in isolation,
this additional component can account only partially for the factor of 
four offset in the derived densities. However, its presence underlines
the substantial extrapolation involved in transforming the observed
number density at large distances from the Plane to a local halo subdwarf
density.  An additional contribution could be due to the choice of
local calibrating subdwarfs for the Gould et al. analysis, which
may be biased towards higher metallicities. 
This would lead to overestimated M$_V$ and distance, and hence an underestimate
of the space density. For a discussion of possible
systematic errors, see Gould et al.'s discussion.  

The kinematics of the very-low-mass halo stars are consistent with those
found for higher mass metal-poor stars.  The strongest
conclusion is that prograde rotating halos are favored. 
If the expanded halo sample that includes lower tangential
velocity stars is included, then kinematics like
those of model A is favored over Layden et al.'s RR Lyrae
kinematics, since the expanded luminosity function is more 
consistent with the model A corrected pure halo sample
(Tables~\ref{table-lf} and~\ref{table-lf-lowvtan}).  
There {\it may} be an excess of sdM at low velocities
over the halo model A predictions.
Since metallicities of $[m/H] \approx -1.2 \pm 0.3$ 
are expected to have some IPII contribution, this 
is not too surprising.  If the two best determined
bins are used, then the excess is only $19\%$.  
Chiba \& Yoshii (1998) find an excess of 
$\sim 20\%$ in the range $-1.4 < [Fe/H] \le -1.0$ based upon 
Hipparcos data for red giant and RR Lyrae stars, in excellent
agreement with our results.

The mass function can be estimated using the
observed luminosity function and a mass-luminosity relation
as follows:
\begin{equation}
\Psi(M) = \Phi(M_I) \frac{dM_I}{dM}
\end{equation}
This is often parameterized as a power-law with the
form
\begin{equation}
\Psi(M)= \frac{dN}{dM} \propto M^{-\alpha}
\end{equation} 
Since there are no direct determinations of mass for 
any sdM or esdM stars, we must rely upon theoretical
mass-luminosity relations.  These must be viewed with caution --
for example, Reid \& Gizis (1997) have shown that, in the case of
the Baraffe et al solar-abundance models, the slope deduced from
nearby-star data for the disk mass function depends on the passband
used in the analysis. Our data do not provide
sufficient range in $M_I$ to estimate the mass function reliably,
even given a good mass-luminosity relationship, but it does
allow us to estimate the relative numbers of sdM and esdM
stars at a mass of $0.2 M_\odot$.   We transform the 
observed pure halo $M_I$ luminosity functions into mass functions
using the Baraffe et al. (1997) models.  We adopt
the $[m/H]= -1.0$ and $[m/H]= -1.5$ interior models 
for the sdM and esdM respectively, and fit a power-law to
the mass function in order to obtain the best estimate of
the numbers. The ratio of sdM to esdM at $0.225 M_\odot$ is 0.38. 
In comparison, we
obtain a ratio of 0.36 if we adopt the Alexander et al. (1997) models, and 0.60
for the D'Antona \& Mazzitelli (1996) tracks --- but the latter
match the observed $V-I,M_I$ main sequence poorly.  
These results are consistent with the Carney et al. (1994)
distribution of metallicities for higher mass stars (mainly G subdwarfs) ---
$41\%$ of their stars in a pure ($V<-220$ km s$^{-1}$)
halo sample have $[m/H]>-1.5$.  Thus, 
the metallicity distribution for the very-low-mass stars
of the halo ($M \approx 0.2 M_\odot$) is apparently the same
as that of stars on the upper main sequence 
($M \approx 0.7 M_\odot$).  Barring a cosmic conspiracy of offsetting
parameters, the most reasonable explanation is that 
the mass function in the halo is independent of metallicity.  
Although our data is inadequate to constrain the mass function
well, we note that the best-fit parameters are  
$\alpha \approx 0.5 \pm 1.6$ for the esdM and $1.3 \pm 0.9$ for the sdM.
These compare with a power-law index of $\alpha = 1.05\pm0.15$  for disk 
stars in the mass range $1.0 > {M \over M_\odot} > 0.1$ (\cite{rg97a}). 

The consistency of the metallicity distribution and kinematics
of the low-mass stars (sdM and esdM) is not surprising, but
nevertheless important.
Studies of the halo, whether using evolved stars
such as RR Lyrae or red giants, or main sequence stars (F,G subdwarfs),
assume that the properties of stars in the narrow mass range from
just below the turnoff to the tip of the asymptotic giant branch
($0.6-0.8 M\odot$, are representative of the halo population as a whole. 
Our results verify one aspect of this assumption for stars near
the bottom of the halo main sequence.

\section{Summary \label{conclusion:halolf}}

We have derived a new estimate of the space density of very low mass
metal-poor stars.  Comparisons with other surveys shows that we are in
agreement with the Dahn {et al.} (1995) survey based upon trigonometric
parallaxes of LHS catalog stars.  This implies that the 
local space density of metal-poor stars is higher than that
predicted by HST observations of more distant stars.

Our work provides measurements of the kinematics and metallicity
distribution of very-low-mass stars.  These measurements are
consistent with those from studies of higher-mass stars in the halo.  
We find that the esdM are 2.5 times as common 
as sdM at $\sim 0.2 M_\odot$.  This is consistent with the 
relative numbers of metal-poor ($[m/H] \le -1.5$) and metal-rich
($[m.H] > -1.5$) G subdwarfs found by Carney et al. (1994).  Note that the
two metallicity scales are comparable because Gizis \& Reid (1997) 
have found agreement between
the Carney et al. metallicities of FGK subdwarf primaries with metallicities
based upon G97.  
The agreement of the metallicity distribution at both low and high mass 
implies that the mass function is not a strong function of
metallicity for ($M \lesssim 0.7 M_\odot$) halo stars.  
This result is supported
by the general similarity of the mass functions derived for the 
two metallicity bins considered; however, those mass functions depend 
upon uncertain model mass-luminosity transformations.

\acknowledgements

We thank Chris Tinney and Mike Irwin for assistance with the plate 
scans and the staff of Palomar Observatory for their capable support. 
We also thank M. Schmidt for useful conversations and 
A. Gould for providing the I-band HST luminosity function.  
J.E.G. is grateful for partial support through 
Greenstein and Kingsley Fellowships as well as NASA grants  
GO-05353.01-93A, GO-05913.01-94A, and GO-06344.01-95A, 
This work is based partly on photographic plates obtained at the Palomar
Observatory 48-inch Oschin Telescope for the Second Palomar
Observatory Sky Survey which was funded by the Eastman Kodak
Company, the National Geographic Society, the Samuel Oschin
Foundation, the Alfred Sloan Foundation, the National Science
Foundation grants AST84-08225, AST87-19465, AST90-23115 and
AST93-18984,  and the National Aeronautics and Space Administration 
grants NGL 05002140 and NAGW 1710.
This research has made use of the Simbad database, operated at
CDS, Strasbourg, France.
  
\appendix
\section{Stars of Particular Interest}

\begin{figure}
\plotone{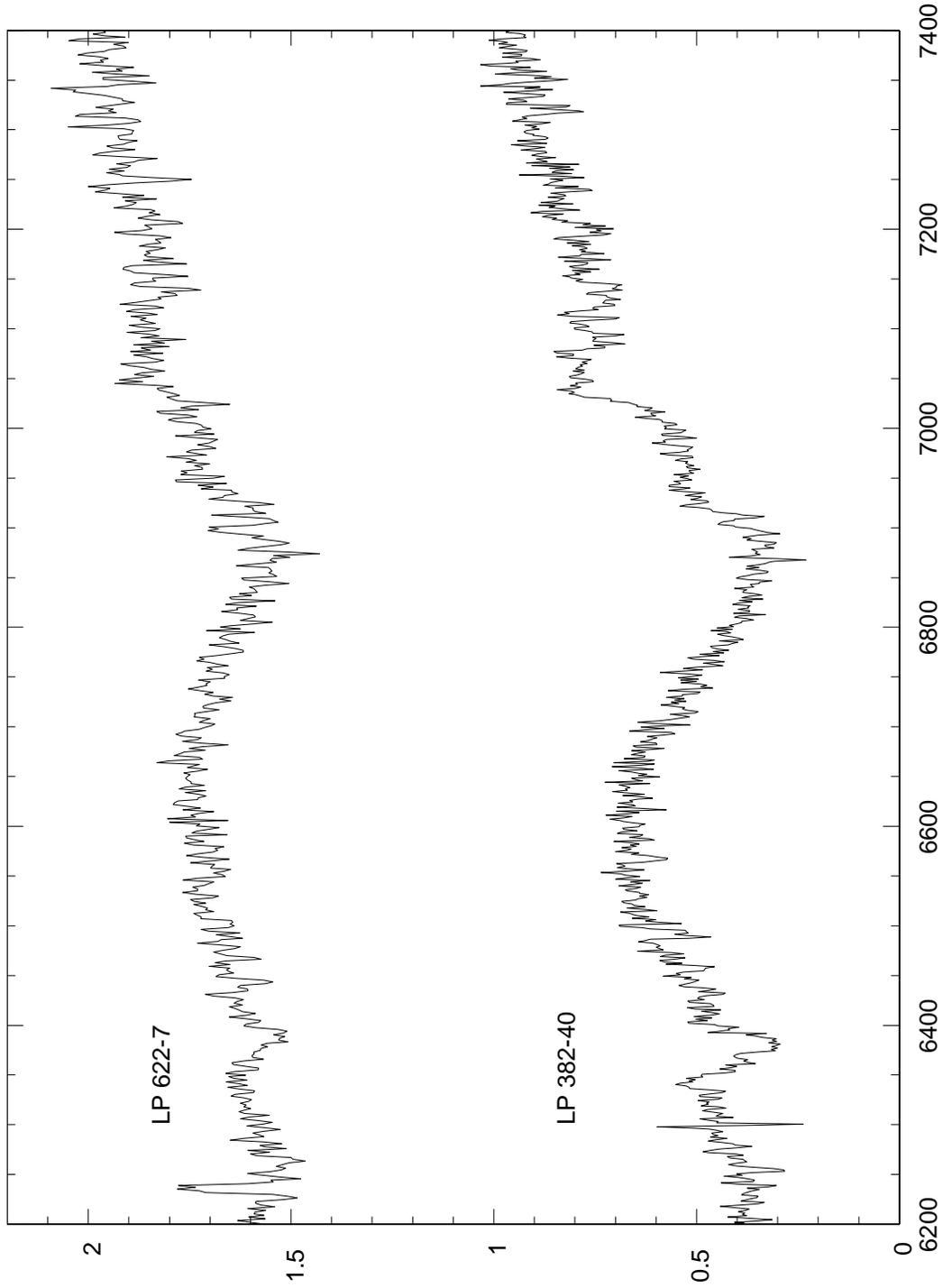}
\caption{The nearby subdwarfs LP 622-7 and LP 382-40.  
The apparent emission lines at the blue end of the spectra
are due to cosmic rays.}
\label{figure-stars1}
\end{figure}

Most of our subdwarfs lie at distances of greater than 200 parsecs,
and therefore are more difficult to study than LHS Catalog subdwarfs.
Three of our stars, however, are close enough that additional
observations may be profitable.  All were previously identified
by Luyten (1979-1980).  The positions (B1950, at the epoch given
in Table~\ref{table-plates}) and other data for these stars are
given in Table~\ref{table-stars}.  Spectra of the subdwarfs LP 622-7 and
LP 382-40 are plotted in Figure~\ref{figure-stars1}.  
LP 622-7 has a proper motion too large to be included
in our complete sample, and was observed at the Palomar 60-in.
in June 1995.  The radial velocity may be unreliable due to a problem
with the calibrating arc lamp exposure.  

\begin{figure}
\plotone{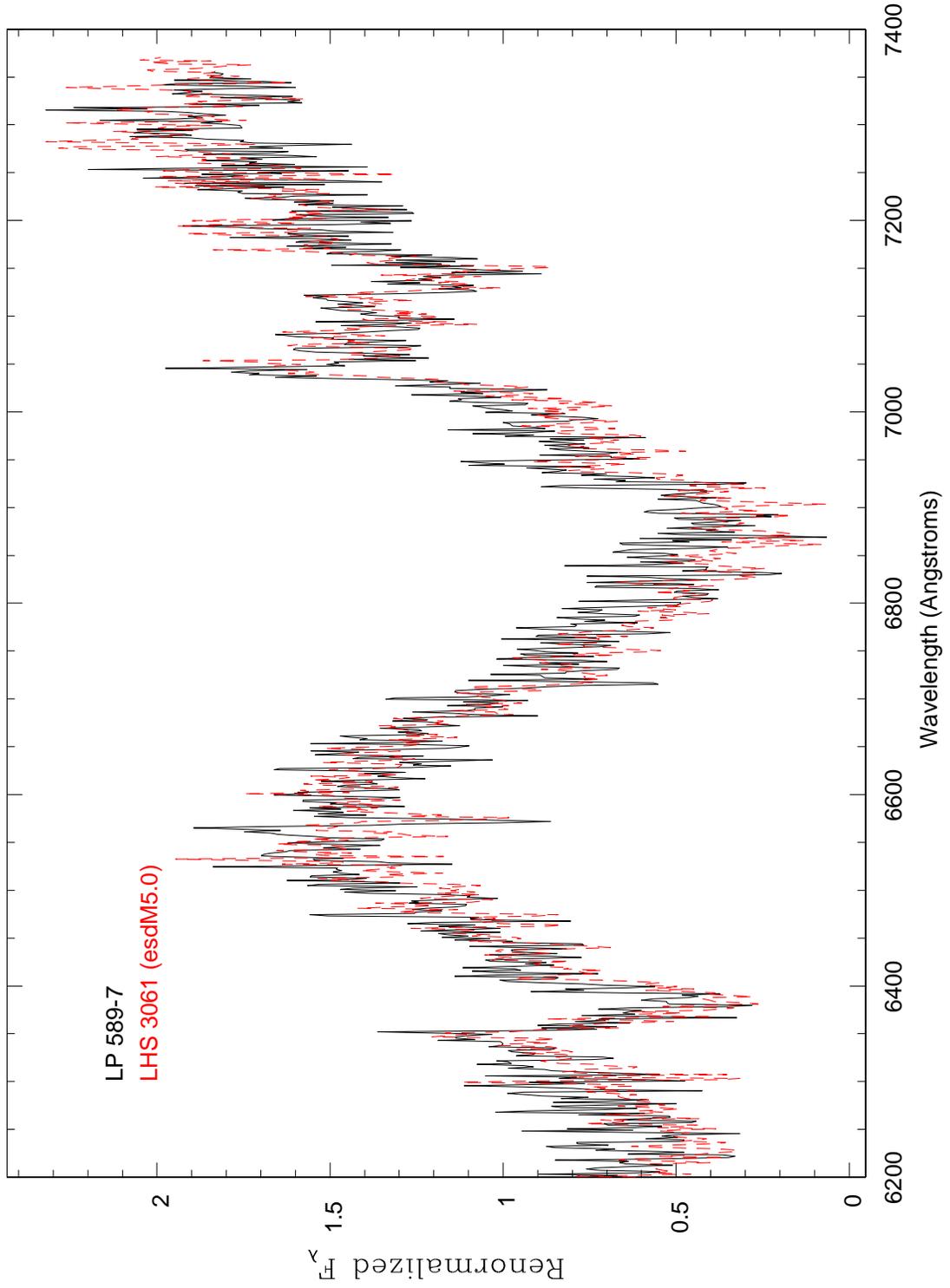}
\caption{The nearby subdwarf LP 589-7 is nearly identical to LHS 3061.
Both are classified as esdM5.0 using the Gizis (1997) system.     
Note the TiO absorption 
in both stars at $7050\AA$, which distinguishes them from 
the other known esdM5.0 star, LHS 205a.}
\label{figure-stars2}
\end{figure}

LP 589-7 is one of the coolest subdwarfs known.  
Its spectral indices are TiO5$=0.68$, CaH1=$0.48$, 
CaH2=$0.27$, and CaH3$=0.50$.  In Figure~\ref{figure-stars2},
we show that this star is extremely similar to the
esdM5.0 (\cite{g97}) star LHS 3061 ($M_I=11.65$, \cite{m92}).   
The conspicuous TiO absorption at $7050 \AA$ distinguishes both
these stars and the slightly cooler LHS 1742a
($M_I=11.69$, esdM5.5) from the other two very cool extreme
subdwarfs LHS 205a ($M_I=11.65$, esdM5.0) and LHS 1826
($M_I$ unknown, esdM6.0; \cite{gr97}).  LP 589-7 is just as close
as LHS 1742a, so it is an excellent candidate for a trigonometric
parallax.  

There is a hint of $H \alpha$ emission in our spectrum of LP 589-7.  
The feature, which may not real but lies at the correct wavelength, 
has an equivalent width of $\sim 1.2 \AA$.    
Two sdM systems (LHS 482 and LHS 2497) 
are known to show emission due to close companions
(\cite{g98}).  Since LP 589-7 is near the hydrogen burning limit, 
any companion must be either a very faint, old white dwarf
(as for the LHS 482 system), a nearly equal luminosity M companion
(as for LHS 2497), or else a brown dwarf (if it proves to be too faint to
contribute detectable light).  Alternatively, LP 589-7 could a 
``young'' metal-poor star with an age of up to a few billion years.  
If an SB2, its estimated distance and tangential velocity should be
increased by up to $40\%$.  
All this is speculative, however, since the $H \alpha$ feature is
weak and may not be real.


\begin{deluxetable}{rrlccclc}
\tablewidth{0pc}
\tablenum{1}
\label{table-plates}
\tablecaption{Survey Fields}
\tablehead{
\colhead{Field}  &
\multicolumn{3}{c}{$\alpha$ (B1950) $\delta$} &
\colhead{Source} &
\colhead{Epoch}  &
\colhead{Plate}  &
\colhead{Epoch} 
}
\tablecolumns{8}
\startdata
829 & 02& 00& +00& UK/PII & 1987.83 & EO852  & 1953.78 \\
831 & 02& 40& +00& UK/PII & 1981.92 & EO1283 & 1954.90 \\
    &   &   &    &        &         & EO1453 & 1955.81 \\ 
832 & 03& 00& +00& UK/PII & 1986.89 & EO363  & 1951.69 \\
    &   &   &    &        &         & EO1453 & 1955.81 \\
262 & 10& 15& +45& POSSII & 1987.34 & EO672  & 1953.12 \\ 
NGP & 13& 04& +29& Palomar& 1976.23 & EO1393 & 1955.29 \\ 
868 & 15& 00& +00& UK/PII & 1987.30 & EO1402 & 1955.30 \\
    &   &   &    &        &         & EO1613 & 1957.32 \\
513 & 15& 00& +25& POSSII & 1987.34 & EO102  & 1950.35 \\ 
    &   &   &    &        &         & EO87   & 1950.30 \\
    &   &   &    &        &         & EO1390 & 1955.29 \\
    &   &   &    &        &         & EO1092 & 1954.49 \\
889 & 22& 00& +00& POSSII & 1990.55 & EO1146 & 1954.57 \\
890 & 22& 20& +00& POSSII & 1990.82 & EO364  & 1951.69 \\
    &   &   &    &        &         & EO1146 & 1954.57 \\
\enddata
\end{deluxetable}

\begin{deluxetable}{crrrr}
\tablewidth{0pc}
\tablenum{2}
\label{table-absmu}
\tablecaption{Relative to Absolute Proper Motions}
\tablehead{
\colhead{Field}  &
\colhead{$\Delta\mu_\alpha$} & 
\colhead{$\Delta\mu_\delta$} &
\colhead{$\sigma_\alpha$} & 
\colhead{$\sigma_\delta$}}
\startdata
 & mas & mas & mas & mas \\
 262 &  2.6 & 5.2 & 13.0 &  13.7 \\
 513 &  4.4 & 4.2 & 12.1 &  11.8 \\
 829 & -3.7 & 3.8 & 14.9 &  13.9 \\
 831 & -2.0 & 2.3 & 17.0 &  16.0 \\
 832 & -4.3 & 3.9 & 13.6 &  12.6 \\
 868 &  3.2 & 3.3 & 13.6 &  12.6 \\
 889 & -0.2 & 4.3 & 10.6 &  10.6 \\
 890 & -0.9 & 3.4 & 11.9 &  11.5 \\
 NGP &  5.0 & 3.9 & \nodata&  \nodata \\
\enddata
\end{deluxetable}

\begin{deluxetable}{lccc}
\tablewidth{0pt}
\tablenum{3}
\label{table-index}
\tablecaption{Spectroscopic Indices}
\tablehead{
\colhead{Band} & \colhead{S1} & \colhead{W} & \colhead{S2}
}
\startdata
TiO 5  & 7042-7046 &7126-7135 & \\
CaH 1  & 6345-6355 &6380-6390 &6410-6420 \\
CaH 2  & 7042-7046 &6814-6846 \\
CaH 3  & 7042-7046 &6960-6990 \\
\enddata
\end{deluxetable}

\begin{deluxetable}{cccc}
\tablewidth{0pc}
\tablenum{4}
\label{table-corr}
\tablecaption{Kinematic Corrections}
\tablehead{
\colhead{Field}  & 
\colhead{Area (sq. deg)} &
\colhead{$m_I^{lim}$} &
\colhead{$1/\chi$} 
}
\startdata
829 & 25.32 & 17.3 & 2.07 \\
831 & 26.49 & 17.3 & 2.15 \\
832 & 25.32 & 17.3 & 2.11 \\
262 & 14.56 & 17.3 & 2.30 \\
NGP & 28.00 & 17.0 & 2.01 \\
868 & 25.29 & 17.3 & 2.11 \\
513 & 25.29 & 17.3 & 2.18 \\
889 & 25.30 & 17.3 & 3.70 \\
890 & 25.26 & 17.3 & 3.56 \\
\enddata
\end{deluxetable}

\begin{deluxetable}{cccccccc}
\tablewidth{0pc}
\tablenum{5}
\label{table-compareuvw}
\tablecaption{Kinematic Comparison}
\tablehead{
\colhead{}  & 
\colhead{Data} & 
\colhead{BC} & 
\colhead{N}  & 
\colhead{L} &
\colhead{CRB} &
\colhead{BSL} &
\colhead{A} 
}
\startdata
$\sigma_U$              & 171& 171 &146 &179 &165 &155 &158  \\ 
Probability (U)         &    & 0.43&0.15&0.55&0.29&0.19&0.31 \\
$\langle V \rangle$     &-234& -222&-244&-240&-252&-264&-233 \\ 
$\sigma_V$              &  76& 82  &82  &84  &73  &76  &85   \\
Probability (V)         &    & 0.37&0.49&0.75&0.22&0.05&0.87 \\
$\sigma_W$              &  90& 77  &86  &97  &94  &105 &86   \\
Probability (W)         &    & 0.38&0.49&0.32&0.38&0.28&0.49 \\
\enddata
\end{deluxetable}

\begin{deluxetable}{rrlrlr}
\tablewidth{0pc}
\tablenum{6}
\label{table-lf}
\tablecaption{Pure Halo Luminosity Functions}
\tablehead{
\colhead{$M_I$} & 
\colhead{$\Phi$ [A]} &
\colhead{$\sigma_\Phi$ [A]} &
\colhead{$\Phi$ [L]} &
\colhead{$\sigma_\Phi$ [L]} &
\colhead{N} }
\startdata
 &\multicolumn{2}{c}{$10^{-5}$ pc$^{-3}$ Mag$^{-1}$}  
 & \multicolumn{2}{c}{$10^{-5}$ pc$^{-3}$ Mag$^{-1}$} \nl
\cutinhead{sdM ($v_{tan}\ge 200$ km s$^{-1}$)}
8.5  & 2.2 & 0.7 & 1.8 & 0.5 & 16 \nl
9.0  & 2.0 & 0.6 & 1.6 & 0.5 & 11 \nl
9.5  & 2.0 & 0.9 & 1.6 & 0.8 & 5  \nl
10.0 & 2.0 & 2.0 & 1.6 & 1.6 & 1  \nl
\cutinhead{esdM ($v_{tan}\ge 200$ km s$^{-1}$)}
9.5  & 4.6 & 1.8 & 3.9 & 1.5 & 10  \nl
10.0 & 3.6 & 1.6 & 2.9 & 1.3 & 5 \nl
10.5 & 3.1 & 2.3 & 2.6 & 2.0 & 2 \nl
\enddata
\end{deluxetable}

\begin{deluxetable}{rrlrlr}
\tablewidth{0pc}
\tablenum{7}
\label{table-lf-lowvtan}
\tablecaption{Expanded Halo Luminosity Functions}
\tablehead{
\colhead{$M_I$} & 
\colhead{$\Phi$ [A]} &
\colhead{$\sigma_\Phi$ [A]} &
\colhead{$\Phi$ [L]} &
\colhead{$\sigma_\Phi$ [L]} &
\colhead{N} }
\startdata
 &\multicolumn{2}{c}{$10^{-5}$ pc$^{-3}$ Mag$^{-1}$}  
 & \multicolumn{2}{c}{$10^{-5}$ pc$^{-3}$ Mag$^{-1}$} \nl
\cutinhead{sdM ($v_{tan}\ge 125$ km s$^{-1}$)}
8.5  & 2.6 & 0.8 & 2.5 & 0.7 & 20 \nl
9.0  & 2.4 & 0.8 & 2.2 & 0.7 & 14 \nl
9.5  & 4.4 & 1.3 & 4.1 & 1.2 & 12  \nl
10.0 & 3.1 & 1.6 & 2.9 & 1.5 & 4  \nl
10.5 & 3.6 & 3.6 & 3.4 & 3.4 & 1  \nl
\cutinhead{esdM ($v_{tan}\ge 100$ km s$^{-1}$)}
9.5  & 6.9 & 1.8 & 6.6 & 1.7 & 20  \nl
10.0 & 2.5 & 1.0 & 2.4 & 1.0 & 6 \nl
10.5 & 4.1 & 2.1 & 3.9 & 2.0 & 4 \nl
11.0 & \nodata & \nodata & \nodata &\nodata & 0 \nl
11.5 & \nodata & \nodata & \nodata &\nodata & 0 \nl
12.0 & 15.\phantom{0} & 15.\phantom{0} & 15.\phantom{0} & 15.\phantom{0} & 1 \nl 
\enddata
\end{deluxetable}

\begin{deluxetable}{lrrrrrrrrrrr}
\tablecaption{Nearby Subdwarfs}
\footnotesize
\tablewidth{0pt}
\tablenum{8}
\label{table-stars}
\tablehead{
\colhead{Name} & \colhead{$\alpha$} & \colhead{$\delta$} & 
\colhead{$\mu_\alpha$} & \colhead{$\mu_\delta$} &
\colhead{$i_P$} & \colhead{$r-i_p$} & \colhead{M$_i$} & 
\colhead{d} & \colhead{$v_{tan}$} & \colhead{$v_{rad}$} 
& \colhead{Sp. Type}
}
\startdata
LP 589-7 & 28.72333 & +1.03647 & -0.12 & -0.33 & 15.90 & 1.38 & 11.77 & 64 & 
106 & -65 & esdM5 \\
LP 382-40 & 223.53375 & +23.83433 & -0.28 & -0.09 & 16.47 & 1.08 & 10.59 & 150 & 210 & -230 & esdM2.5/3 \\
LP 622-7 & 227.18958 & -1.02281 & +0.20 & -0.43 & 13.87 & 0.87 & 9.69 & 69 & 155 & 300 & esdK7/M0 \\
\enddata
\end{deluxetable}

\end{document}